\documentclass[12pt]{article}
\usepackage{amsmath}
\usepackage{graphicx}%
\usepackage{amsfonts}%
\usepackage{amssymb}
\usepackage{enumerate}
\usepackage{natbib}
\usepackage{url} 
\usepackage{color}
\usepackage{multicol}
\usepackage{multirow}
\usepackage{booktabs}
\usepackage{makecell}
\usepackage{tikz}
\usepackage{algorithm}
\usepackage{algpseudocode}
\usetikzlibrary{positioning, arrows.meta}
\usepackage{color,soul}
\usepackage{hyperref}

\usepackage{float}

\usepackage{xcolor}
\definecolor{pastelred}{RGB}{235,170,170}
\definecolor{pastelblue}{RGB}{165,190,220}
\definecolor{pastelgreen}{RGB}{190,215,170}

\usepackage{tikz}
\usetikzlibrary{calc}
\usetikzlibrary{positioning}

\newcommand{\blind}{1}

\newcommand{\btheta}{\boldsymbol{\theta}}

\begin{document}

\def\spacingset#1{\renewcommand{\baselinestretch}%
{#1}\small\normalsize} \spacingset{1}

\if1\blind
{
  \title{\bf 
  Modeling Animal Communication Using Multivariate Hawkes Processes with Additive Excitation and Multiplicative Inhibition
  }
  \author{
    Bokgyeong Kang\\
    Department of Statistics, Dongguk University,\\
    \\
    Erin M. Schliep\\
    Department of Statistics, North Carolina State University,\\
    \\
    Alan E. Gelfand\\
    Department of Statistical Science, Duke University,\\
    \\
    Ariana Strandburg-Peshkin\\
    Department for the Ecology of Animal Societies,\\ Max Planck Institute of Animal Behavior,\\
    Department of Biology, University of Konstanz,\\
    \\
    Robert S. Schick\\
    Southall Environmental Associates, Inc.
  }
  \maketitle
} \fi

\if0\blind
{
  \bigskip
  \bigskip
  \bigskip
  \begin{center}
    {\LARGE\bf Modeling Animal Communication Using Multivariate Hawkes Processes with Additive Excitation and Multiplicative Inhibition}
\end{center}
  \medskip
} \fi

\begin{abstract}
Animal acoustic communication often exhibits temporal dependence, with calls triggering or suppressing subsequent calls within and across call types, individuals, or species. While Hawkes processes provide a natural framework for modeling excitation, incorporating inhibition in multivariate settings can raise identifiability issues and complicate parameter interpretation.
We propose a flexible class of multivariate Hawkes processes that combines additive excitation with multiplicative inhibition. This formulation preserves the branching process interpretation of excitation while reducing confounding between excitation and inhibition, and allows direct quantification of background and excitation contributions to the event rate. Bayesian inference is conducted via Markov chain Monte Carlo, and model adequacy is assessed using the random time change theorem.
The proposed methodology is evaluated through simulation and applied to two acoustic communication datasets: group-living meerkats, for which we analyze three selected call types with distinct behavioral roles, and a two-species baleen whale dataset involving humpback and North Atlantic right whales.
The meerkat analysis reveals significant within- and cross-type excitation with cross-type inhibition, whereas the whale data show evidence primarily of within-species excitation.
\end{abstract}

\noindent%
{\it Keywords: Event time sequences; hierarchical models; identifiability; meerkat calling; random time change theorem; temporal point patterns; whale calling} 

\spacingset{1.6}



\section{Introduction}

An important biological event time process concerns animal communication \citep{bradburyPrinciplesAnimalCommunication1998, seyfarthSignalersReceiversAnimal2003} where the event sequence consists of a collection of calls made by one or multiple individuals over time \citep{kershenbaumAcousticSequencesNonhuman2016}. When collecting and analyzing this type of data, the objective is often to identify the relationship between characteristics of calls or sequences of calls and observable patterns of animal behavior. Such analyses can reveal important insights into the function of vocalizations within different animal species, with implications for the evolution of communication systems across taxa. Beyond responses to individual calls, much information can often be gleaned from the temporal structure of vocal sequences.
Particular acoustic communication patterns could include excitation or inhibition, where calls encourage more (excitation) or less (inhibition) calling, either within or across individuals. For example, so-called \emph{contact} calls have been documented across many taxa \citep{rendall2000proximate}, and often show call-response patterns where a call from an individual leads to increased calling activity from neighboring individuals \citep{engesser2022collective,miller2009antiphonal}. Similarly, \emph{alarm} calls may signify the presence of predators \citep{manserSuricateAlarmCalls2002} and could lead to subsequent inhibition of calling to avoid detection. In some cases, both excitation and inhibition may be at play, but operate at different time scales. For example, meerkats avoid producing sunning calls immediately after hearing their conspecifics call, thus avoiding overlap, but increase their call rates over longer timescales. Both excitation and inhibition may also operate across call types. For example, aggressive vocalizations could lead to submissive calling in response (cross-excitation). The impacts of calling behavior could depend further on the acoustic structure of the call or the individual making the call. Overall, the structure of vocal sequences within and across individuals has the potential to encode a great deal of information, and understanding the processes driving this structure---including both excitation and inhibition---can yield important insights into animal communication systems. 

Observed call sequences can typically be represented as temporal or spatiotemporal event sequences, and are often modeled using point processes. Hawkes processes have emerged as a popular temporal point process model specification to capture excitement in event sequences.  Dating to \citet{Hawkes1971}, the univariate version of these models introduces a conditional intensity which adds a positive triggering component to a background intensity to reflect previous event history.  Under this model, previous events and, in particular, more recent previous events elevate the intensity to encourage more events at the current time.  Motivating examples include seismology, epidemiology, crime modeling, mathematical finance, and events in social media \citep{Ogata1998,Meyer2012,mohler2011self,hawkes2018hawkes,rizoiu2017hawkes}.

Multivariate extensions of Hawkes process models can be defined for a collection of interacting processes where the overall conditional process intensity arises as the sum of the individual conditional intensities. Multivariate Hawkes processes have experienced an increase in utilization.  Applications consider interacting event sequences that arise in areas such as finance, social dynamics, neurosciences, and ecological science \citep{embrechts2011multivariate,zhou2013learning,reynaud2013inference,Nicvert2024,Kang2025whaleHawkes}. Given their multivariate nature, these models incorporate both self-excitation (events increase the rate of future events of the same type) and cross-excitation (events increase the rate of future events of different types).

More recent attention has been focused on incorporating inhibition into these processes \citep{Mei2017,Bonnet2023,Sulem2024,Deutsch2025}.
As opposed to excitement processes, with inhibition processes, previous events diminish the intensity to encourage fewer events at the current time. Under an additive model specification, the background intensity as well as the excitement component are positive while the inhibition component is negative. As a result, a link function is often introduced in order to maintain a non-negative intensity. General models supplementing a background process with an excitement or an inhibition component are referred to as evolutionary processes; various link functions and their behavior are discussed in \citealt{White2021}
Despite these developments, jointly modeling excitation and inhibition remains challenging. Given only a sequence of events, there are no labels or marks to identify which events are exciting or are inhibiting future events. However, the model for the intensity needs to enable the incidence of excitement and inhibition to attempt to explain the event sequence.

The contribution of this work is the development of novel forms for the conditional intensity function that allow for the identification of both excitement and inhibition. 
In the multivariate setting, our model specification allows for self- and cross-excitation as well as self- and cross-inhibition. Our conditional intensity model incorporates additive triggering functions for excitement with nonnegative multiplicative scaling functions for inhibition.  Our work is motivated by \citet{Olinde2020}, who consider this class of models in the univariate setting. However, as we outline below, their work doesn't fully appreciate the identifiability challenges arising with these models. The explicit benefits of our specification are detailed in Section~\ref{sec:motivation}.  
We propose a very general and flexible class of multivariate excitement and inhibition models that yield clear interpretation. We present computational strategies for sampling realizations from these processes that result in well-behaved model inference. Further, we offer model comparison and assessment through the random time change theorem \citep[RTCT;][]{Brown1988}, viewed within a Bayesian framework.  

We demonstrate model performance through simulation as well as the analysis of two real datasets: (i) a single species dataset comprised of group-living meerkats (\textit{Surricatta suricata}) making three different types of calls, each of which is known to have different behavioral consequences \citep{Demartsev2024}; and (ii) a multi-species dataset of two different baleen whales (humpback whales, \textit{Megaptera novaengliae}, and North Atlantic right whales \textit{Eubalaena glacialis}), which forage in similar spatial locations and who each have diverse acoustic repertoires \citep{paynemcvay1971,clarkAcousticRepertoireSouthern1982}.

The format of the paper is as follows.  In Section 2 we provide a brief review of Hawkes processes, univariate and multivariate, including those with inhibition.  Section 3 elaborates our model specification for a multivariate Hawkes process with additive excitement and multivariate inhibition.  Section 4 provides full model specification in a Bayesian framework including prior specification, model fitting, and model assessment.  Section 5 supplies the results of several simulation experiments which serve as proofs of concept.  Section 6 presents the analysis of both the meerkat data and the whale data.  In Section 7, we conclude with a summary and several avenues for future work.

\section{Background on Hawkes Processes}
\label{sec:background}

This section reviews the basic formulation of Hawkes processes. We introduce the univariate and multivariate models and summarize existing approaches for incorporating inhibitory effects, which inform the modeling choices discussed in the next section.

\subsection{Univariate Hawkes Process}

We first review the univariate Hawkes process. Let $\mathcal{T} = \{ t_i : i = 1, \dots, n \}$ denote the observed sequence of event times, where $t_i \in (0, T]$ and $t_i < t_{i+1}$. Let $\mathcal{H}_t$ denote the history of the process up to time $t$, defined as $\mathcal{H}_t = \{ t_i : t_i < t \}$. A temporal point process is characterized by its conditional intensity function, defined as
\begin{align*}
    \lambda(t \mid \mathcal{H}_t)
    &=
    \lim_{\Delta t \to 0}
    \frac{
    \mathbb{E}\left\{
    N\bigl((t, t + \Delta t]\bigr)
    \mid \mathcal{H}_t
    \right\}
    }{\Delta t},
\end{align*}
where $N(A)$ denotes the number of events occurring in a measurable set $A \subseteq (0, T]$.

The conditional intensity function of a univariate Hawkes process at time $t$ is specified as
\begin{align}
    \lambda(t \mid \mathcal{H}_t; \boldsymbol{\theta})
    &=
    \mu(t; \boldsymbol{\theta}_{\mu})
    +
    \sum_{i : t_i < t}
    \alpha \, g(t - t_i; \boldsymbol{\theta}_g), \label{eq:uniHawkes}
    \end{align}
where $\mu(\cdot)$ denotes the background (or baseline) intensity governing events that occur independently of the past history. This component may accommodate temporal inhomogeneity such as seasonality or long-term trends.
The function $g(\cdot)$ is a nonnegative triggering kernel satisfying $g(x) \ge 0$ for $x \ge 0$ and $\int_{0}^{\infty} g(x)\,dx = 1$, which determines how the influence of a past event decays over time. Common choices for $g(\cdot)$ include exponential kernels, as well as other parametric forms such as power-law and Gaussian kernels. The parameter $\alpha$ controls the overall magnitude of this influence. In the classical Hawkes process, the restriction $\alpha \ge 0$ is imposed, so that past events can only increase the future event rate, yielding a self-exciting process.

The model parameters are given by $\boldsymbol{\theta} = (\boldsymbol{\theta}_{\mu}, \alpha, \boldsymbol{\theta}_g)$,
where $\boldsymbol{\theta}_{\mu}$ and $\boldsymbol{\theta}_g$ collect the parameters associated with $\mu(\cdot)$ and $g(\cdot)$, respectively. A variety of estimation approaches have been proposed in the literature, including maximum likelihood and Bayesian methods \citep{Veen2008,rasmussen2013bayesian,Ross2021}.

\subsection{Multivariate Hawkes Process}

The univariate Hawkes model in \eqref{eq:uniHawkes} extends naturally to a $K$-dimensional multivariate Hawkes process that accommodates both self-excitation and cross-excitation among event types. The observed data now consist of marked event times, represented as $\mathcal{T} = \{ (t_i, m_i) : i = 1, \dots, n \}$, where $t_i \in (0, T]$ denotes the occurrence time of the $i$th event and $m_i \in \{1, \dots, K\}$ denotes its associated mark or event type. The history of the process up to (but not including) time $t$ is given by $\mathcal{H}_t = \{ (t_i, m_i) : t_i < t \}$.

The conditional intensity function for mark $k$ at time $t$ is given by
\begin{align}
    \lambda_k(t \mid \mathcal{H}_t; \boldsymbol{\theta})
    &=
    \mu_k(t; \boldsymbol{\theta}_{\mu})
    +
    \sum_{\ell = 1}^K
    \sum_{\substack{i : t_i < t, \\ m_i = \ell}}
    \alpha_{\ell,k}\,
    g_{\ell,k}(t - t_i; \boldsymbol{\theta}_g). \label{eq:additive}
\end{align}
Here, $\mu_k(\cdot)$ denotes the mark-specific background intensity. Past events of all marks may contribute to the current intensity for mark $k$, with the contribution from events of mark $\ell$ governed by the interaction parameter $\alpha_{\ell,k}$ and the triggering kernel $g_{\ell,k}(\cdot)$. This formulation allows for both self-excitation ($\ell = k$) and cross-excitation ($\ell \ne k$) across marks. 

The model parameters are given by $\boldsymbol{\theta} = (\boldsymbol{\theta}_{\mu}, \boldsymbol{\alpha}, \boldsymbol{\theta}_g)$, where $\boldsymbol{\alpha} = \{ \alpha_{\ell,k} : \ell, k = 1, \dots, K \}$ collects the interaction parameters across marks. The overall conditional intensity of events at time $t$ is obtained by summing the mark-specific intensities, $\lambda(t \mid \mathcal{H}_t; \boldsymbol{\theta}) = \sum_{k=1}^K \lambda_k(t \mid \mathcal{H}_t; \boldsymbol{\theta})$.

\subsection{Hawkes Process with Inhibition}
\label{sec:HawkesInhibition}

Several recent extensions of the multivariate Hawkes process allow for inhibitory effects by permitting $\alpha_{\ell,k} < 0$ \citep{Mei2017,Bonnet2023,Sulem2024,Deutsch2025}. In this case, an event of mark $\ell$ reduces the future intensity of mark $k$, thereby decreasing the likelihood of subsequent events of type $k$. Because negative interaction parameters may lead to negative intensities, a link function is introduced to enforce nonnegativity. This yields the intensity function
\begin{align}
    \lambda_k(t \mid \mathcal{H}_t; \boldsymbol{\theta})
    &=
    \phi\!\left(
    \mu_k(t; \boldsymbol{\theta}_{\mu})
    +
    \sum_{\ell = 1}^K
    \sum_{\substack{i : t_i < t, \\ m_i = \ell}}
    \alpha_{\ell,k}\,
    g_{\ell,k}(t - t_i; \boldsymbol{\theta}_g)
    \right),
\end{align}
where $\alpha_{\ell,k} \in \mathbb{R}$ and the link function $\phi(\cdot)$ ensures that the resulting intensity remains nonnegative. Common choices for $\phi(\cdot)$ include the rectified linear unit (ReLU), $\phi(x) = \max(x, 0)$, the sigmoid function,
$\phi(x) = (1 + e^{-x})^{-1}$, and the softplus function, $\phi(x) = \log(1 + e^{x})$, among others.

Unlike additive inhibition models, in which excitation and inhibition enter the intensity additively, \citet{Olinde2020} introduce inhibition through a
multiplicative component modulating a self-exciting intensity. They propose a univariate model, referred to as the \emph{self-limiting} Hawkes process, with conditional intensity given by
\begin{align}
    \lambda(t \mid \mathcal{H}_t; \boldsymbol{\theta})
    &=
    \left(
    \mu
    +
    \sum_{i : t_i < t}
    \alpha \, g(t - t_i; \boldsymbol{\theta}_g)
    \right)
    \exp\{-\gamma \, N(\phi, t)\},
\end{align}
where $\gamma > 0$ controls the strength of inhibition. Larger values of $\gamma$ correspond to stronger inhibition of future events. Here, $N(\phi, t)$ denotes the number of events occurring in the interval $[t - \phi, t)$, so that a higher frequency of recent events leads to a greater reduction in the current intensity. The background intensity $\mu$ is assumed to be constant. The model parameters are given by $\boldsymbol{\theta} = (\mu, \alpha, \boldsymbol{\theta}_g, \gamma, \phi)$.

This formulation has been shown to perform well in regimes where the inhibitory effect is moderate. Our simulation results suggest, however, that when both excitation and inhibition are strong, the corresponding parameters may be difficult to identify separately. 
Details are provided in Supplementary Section~S1. 
Related concerns have also been raised in \citet{Olinde2020} for regimes with $\alpha > 1$, which in standard Hawkes processes are associated with unstable or explosive growth of the event intensity. In the presence of multiplicative inhibition, large excitation parameters $\alpha$ interact directly with the inhibition parameter $\gamma$ through the damping term $\exp\{-\gamma N(\phi,t)\}$, reflecting a competition between excitatory growth and inhibitory suppression that can further exacerbate identifiability issues.
We further note that the inhibition term $\exp\{-\gamma \, N(\phi, t)\}$ is piecewise constant in time, since $N(\phi, t)$ is a discrete-valued counting process.  These comments motivate an alternative parameterization, presented in the following section, that aims to alleviate identifiability challenges while providing a continuous representation of inhibition.

\section{Multivariate Hawkes Processes with Additive Excitation and Multiplicative Inhibition}
\label{sec:model}


This section introduces a multivariate Hawkes process with additive excitation and multiplicative inhibition. We present the model specification and briefly discuss the motivation for this structure.

\subsection{Model Specification}

Building on the framework of \citet{Olinde2020}, we propose a model that accommodates both excitatory and inhibitory interactions across multiple mark types. For mark $k = 1, \ldots, K$, we define the conditional intensity function as
\begin{align}
    \lambda_k(t \mid \mathcal{H}_t; \boldsymbol{\theta})
    &=
    \left(
    \mu_k(t; \btheta_{\mu})
    +
    \sum_{\ell = 1}^K
    \sum_{\substack{i : t_i < t, \\ m_i = \ell}}
    \alpha_{\ell,k}\,
    g_{\ell,k}(t - t_i; \boldsymbol{\theta}_g)
    \right)
    \exp\!\left\{
    -
    \sum_{\ell = 1}^K
    \sum_{\substack{i : t_i < t, \\ m_i = \ell}}
    \gamma_{\ell,k}\,
    h_{\ell,k}(t - t_i; \boldsymbol{\theta}_h)
    \right\},
    \nonumber
\end{align}
where the background intensity is modeled as $\mu_k(t) = \mathbf{x}(t)\boldsymbol{\beta}_k$, with $\mathbf{x}(t)$ denoting a row vector of covariates observed at time $t$.

For the excitation kernels, we employ the exponential form $g_{\ell,k}(x)= \eta_{\ell,k}^{-1} \exp(-x / \eta_{\ell,k})$ for $x > 0$. This choice assigns greater weight to more recent events, with the influence of past events decaying exponentially over time. The parameter $\eta_{\ell,k} > 0$ governs the rate of this decay and, in general, may vary across marks, allowing different mark types to exhibit distinct temporal decay patterns. For parsimony, we further assume that $\eta_{\ell,k} = \eta_{\ell}$, so that the temporal range of the excitatory influence of mark $\ell$ is shared across all receiving marks $k$.

For the inhibition kernels, we adopt an exponential specification, $h_{\ell,k}(x)= \exp(-x / \phi_{\ell,k})$ for $x > 0$. Under this formulation, inhibitory effects are strongest immediately following
an event and decay gradually as the time lag increases. The parameter $\phi_{\ell,k} > 0$ controls the rate of this decay and, in general, may vary across marks, allowing different mark types to exhibit distinct temporal inhibition behaviors. For parsimony, we further impose $\phi_{\ell,k} = \phi_{\ell}$, implying that the temporal range of inhibition associated with mark $\ell$ is common across all marks $k$.

As discussed in Section~\ref{sec:HawkesInhibition}, excitation and inhibition parameters may be difficult to identify separately when both effects are strong, that is, when $\alpha_{\ell,k} > 0$ and $\gamma_{\ell,k} > 0$ take sufficiently large values. To address this identifiability challenge, we impose the constraint
\begin{align}
    \alpha_{\ell,k} \ge 0, \quad \gamma_{\ell,k} \ge 0, \quad
    \text{and} \quad \alpha_{\ell,k}\,\gamma_{\ell,k} = 0,
    \label{eq:constraint}
\end{align}
which restricts each ordered pair $(\ell, k)$ to a \emph{single} type of interaction,
either excitation or inhibition. Under this constraint, a mark type $\ell$ cannot simultaneously increase and decrease the intensity of mark type $k$. The interactions are directional, allowing for asymmetric relationships; for example, mark $\ell$ may excite mark $k$, while mark $k$ inhibits mark $\ell$.
We note that a similar exclusivity is implicitly assumed in additive-additive
formulations, where excitation and inhibition are represented through a single signed interaction parameter. In such models, a given ordered pair $(\ell,k)$ is interpreted as either excitatory or inhibitory, depending on the sign of the interaction coefficient, but not both simultaneously. From this perspective, the constraint in \eqref{eq:constraint} formalizes an assumption that is conceptually consistent with existing additive-additive approaches, while providing a clearer separation of excitatory and inhibitory effects. To enforce the constraint, we introduce latent indicator variables $I_{\alpha_{\ell,k}}$ and $I_{\gamma_{\ell,k}}$ and reparameterize the interaction parameters as $\alpha_{\ell,k} = \alpha^{*}_{\ell,k}\, I_{\alpha_{\ell,k}}$ and $\gamma_{\ell,k} = \gamma^{*}_{\ell,k}\, I_{\gamma_{\ell,k}}$. The pair $(I_{\alpha_{\ell,k}}, I_{\gamma_{\ell,k}})$ is restricted to take values in $\{(1,0),\ (0,1),\ (0,0)\}$, thereby ensuring that excitation and inhibition cannot occur simultaneously for any ordered pair $(\ell,k)$.

\subsection{Motivation for the Additive–Multiplicative Parameterization}
\label{sec:motivation}

Our formulation combines excitation additively and inhibition multiplicatively, which offers several methodological advantages over approaches that incorporate both effects additively. Below, we outline the key benefits of this interaction structure.

\subsubsection{Improved Interpretability}
Under additive-additive formulations, inhibitory effects enter additively and compete with both the background and excitation terms, potentially offsetting their combined contribution to the intensity. In contrast, multiplicative inhibition modulates the overall event rate by scaling both background and excitation components proportionally, which yields a clearer separation and interpretation of their respective contributions. Such behavior aligns with many biological and behavioral systems, in which inhibitory effects modulate overall activity levels instead of directly offsetting excitatory influences.

Moreover, additive–additive models typically rely on a link function to enforce
nonnegativity of the intensity, which obscures the \textit{direct} interpretation of model parameters available in standard Hawkes processes. For example, under such formulations, excitation parameters no longer admit a direct branching-process interpretation (see Section~\ref{sec:likelihood} for details), and decay parameters cannot be interpreted straightforwardly in terms of the average persistence of excitation. In contrast, the proposed additive–multiplicative specification preserves these interpretations. Although inhibition enters the model multiplicatively, it acts as a modulation of the overall intensity and does not distort the underlying additive structure of the excitation component. As a result, the excitation magnitude can still be understood in terms of offspring contributions, and the decay parameters retain their usual interpretation as governing the temporal persistence of excitation.

\subsubsection{Separation of Background and Excitation under Inhibition}
In the additive-additive specification in \eqref{eq:additive}, excitation and
inhibition are encoded through a single signed interaction parameter $\alpha_{\ell,k}$, with $\alpha_{\ell,k} > 0$ corresponding to excitation and
$\alpha_{\ell,k} < 0$ to inhibition, and a link function $\phi(\cdot)$ is required to ensure nonnegativity of the resulting intensity. Under this formulation, the contributions of the background intensity and the excitation mechanism are intrinsically confounded, making it impossible to separately identify the expected number of background events and the expected number of excitation-driven events.

In contrast, the proposed additive--multiplicative parameterization yields a clear separation between background and excitation components, each modulated by an inhibition term. Specifically, the conditional intensity for mark $k$ can be expressed as
\begin{align*}
\lambda_k(t)
&=
\bigl\{ \mu_k(t) + G_k(t) \bigr\} \, H_k(t)
=
\mu_k(t)\, H_k(t)
\;+\;
G_k(t)\, H_k(t),
\end{align*}
where $G_k(t) = \sum_{\ell} \sum_{i} \alpha_{\ell,k} \, g_{\ell,k}(t - t_i)$ denotes the excitation component, and $H_k(t) = \exp\!\left\{ - \sum_{\ell} \sum_{i} \gamma_{\ell,k} \, h_{\ell,k}(t - t_i) \right\}$ denotes the multiplicative inhibition factor. Under this decomposition, the first term corresponds to background events modulated by inhibition, while the second term represents excitation-driven events, also modulated by the same inhibition mechanism. 

This decomposition admits a natural interpretation in terms of \textit{compensators}. For a point process with conditional intensity $\lambda(t \mid \mathcal{H}_t; \boldsymbol{\theta})$, the compensator $\Lambda(t \mid \mathcal{H}_t; \boldsymbol{\theta}) = \int_0^t \lambda(u \mid \mathcal{H}_u; \boldsymbol{\theta})\,du$ represents the expected number of events over $(0,t]$ and is a random quantity, as it depends on the observed history. Under the proposed model, the additive--multiplicative structure implies that the overall compensator can be decomposed into subcompensators associated with the background and excitation components. Accordingly, the expected numbers of background and excitation-driven events of mark $k$ over the interval $(0,T]$ are given by
\begin{align}
    \mathbb{E}\{ N_{k,\text{bg}}(0,T] \}
    &=
    \int_0^T \mu_k(t)\, H_k(t)\, dt,
    \nonumber
    \\ 
    \mathbb{E}\{ N_{k,\text{exc}}(0,T] \}
    &=
    \int_0^T G_k(t)\, H_k(t)\, dt, 
    \nonumber
\end{align}
where $N_{k,\mathrm{bg}}(0,T]$ and $N_{k,\mathrm{exc}}(0,T]$ denote the numbers of background and excitation-driven events of mark $k$, respectively. From a Bayesian perspective, posterior samples of the model parameters induce posterior distributions over these compensators, enabling direct uncertainty quantification for the expected contributions of background activity and excitation.

\subsubsection{Enhanced Computational Efficiency}
Under the proposed additive-multiplicative formulation, the intensity is linear
in the background component and linear in each excitation component, conditional
on the multiplicative inhibition factor. As a result, parameter updates can be carried out using only the relevant component of the intensity. For example, updating background parameters involves only the background-inhibition term $\mu_k(t)\, H_k(t)$, whereas updating excitation parameters requires evaluating only the excitation-inhibition term $G_k(t)\, H_k(t)$. This separation reduces the computational cost of individual MCMC updates and can lead to improved mixing behavior. In contrast, additive-additive formulations require evaluating the full intensity at each parameter update, resulting in increased computational burden and typically slower mixing.

\section{Bayesian Inference}
\label{sec:inference}

In this section, we describe the Bayesian inference framework for the proposed model, including the likelihood formulation, prior specification, posterior computation, and model assessment procedures.

\subsection{Likelihood}
\label{sec:likelihood}

The likelihood function of the model \citep{White2021} is given by
\begin{align}
L(\boldsymbol{\theta} \mid \mathcal{T})
&=
\exp\!\left\{
- \sum_{k=1}^K
\int_{0}^T
\lambda_k(t \mid \mathcal{H}_{t}; \boldsymbol{\theta}) \, dt
\right\}
\prod_{i=1}^n
\lambda_{m_i}(t_i \mid \mathcal{H}_{t_i}; \boldsymbol{\theta}), \label{eq:lik}
\end{align}
where $\mathcal{T} = \{(t_i, m_i) : i = 1, \dots, n\}$ denotes the observed marked
event times. The model parameters are collected in $\boldsymbol{\theta} = (\boldsymbol{\beta}, \boldsymbol{\alpha}, \boldsymbol{\eta}, \boldsymbol{\gamma}, \boldsymbol{\phi})$, with $\boldsymbol{\beta} = \{ \boldsymbol{\beta}_{k} : k = 1,\dots,K \}$, $\boldsymbol{\eta} = \{ \eta_{\ell} : \ell = 1,\dots,K \}$, $\boldsymbol{\gamma} = \{ \gamma_{\ell,k} : \ell, k = 1,\dots,K \}$, and $\boldsymbol{\phi} = \{ \phi_{\ell} : \ell = 1,\dots,K \}$.

Estimation for this model can be viewed as an incomplete-data problem.
Following \citet{Veen2008}, we introduce latent branching variables
$z_i$ that indicate whether the event at time $t_i$ is a background event or is
triggered by a previous event, namely,
\begin{align*}
    z_i
    =
    \begin{cases}
    0, & \text{if the event at } t_i \text{ is a background event}, \\
    j, & \text{if the event at } t_i \text{ is excited by a past event at } t_j,
    \end{cases}
\end{align*}
for $i = 1, \dots, n$, where $j \in \{ r : t_r < t_i \}$.
Let $\boldsymbol{z} = (z_1, \dots, z_n)$ denote the collection of latent
branching indicators.
Conditional on $\boldsymbol{z}$, the observed events can be partitioned into
$n+1$ mutually exclusive sets $\mathcal{T}_0, \mathcal{T}_1, \dots, \mathcal{T}_n$,
where $\mathcal{T}_j = \{ (t_i, m_i) : z_i = j \}$ for $j = 0, \dots, n$.
Here, $\mathcal{T}_0$ corresponds to background events, while $\mathcal{T}_j$
for $j > 0$ consists of events excited by the event occurring at time $t_j$.
The union of these sets recovers the full set of observed events.

Under the proposed additive--multiplicative formulation, the latent branching
structure induces a convenient decomposition of the likelihood.
Specifically, conditional on the branching indicators $\boldsymbol{z}$,
the set $\mathcal{T}_0$ of background events follows a nonhomogeneous Poisson
process (NHPP) with intensity $\mu_k(t)\,H_k(t)$ across marks $k$, whereas each
$\mathcal{T}_j$ for $j>0$ corresponds to an NHPP with intensity
$G_k(t)\,H_k(t)$ across marks.
This structure allows the complete-data likelihood to be expressed as
\begin{align}
L(\boldsymbol{\theta} \mid \mathcal{T}, \boldsymbol{z})
&\propto
\exp\!\left\{
- \sum_{k=1}^K
\int_0^T
\lambda_k(t)\, dt
\right\}
\prod_{i=1}^n
\bigl[
\mu_{m_i}(t_i)\, H_{m_i}(t_i)
\bigr]^{\mathbb{I}(z_i = 0)}
\prod_{i=1}^n
\bigl[
G_{m_i}(t_i)\, H_{m_i}(t_i)
\bigr]^{\mathbb{I}(z_i > 0)},
\nonumber
\end{align}
where $\mathbb{I}(\cdot)$ denotes the indicator function. Such a factorization is a direct consequence of the multiplicative inhibition structure. In contrast, under additive--additive formulations with a link function, excitation and inhibition enter the intensity through a nonlinear transformation, which prevents a comparable decomposition of the likelihood into separate background and excitation contributions.

The availability of this decomposition enables efficient data augmentation via
the latent branching variables, which, in turn, facilitates parameter updates
based on lower-dimensional components of the likelihood. As a result, introducing the latent branching structure can improve mixing and convergence of the Markov chain Monte Carlo algorithm. Posterior inference is carried out using a Metropolis--Hastings-within-Gibbs sampler, and numerical integration is employed to evaluate the integrals appearing in the likelihood.

\subsection{Prior Specification}

The background regression coefficients $\boldsymbol{\beta}_k$ are assigned
independent standard normal priors, $\boldsymbol{\beta}_k \sim \mathrm{N}(\mathbf{0}, 10 \mathbf{I})$. To encourage sparsity and enhance interpretability of the interaction structure,
we place spike-and-slab priors on the excitation and inhibition parameters
$\alpha_{\ell,k}$ and $\gamma_{\ell,k}$.
Specifically, each interaction coefficient is modeled as a mixture of a point
mass at zero and a log-normal distribution,
\[
\log \alpha_{\ell,k} \sim (1 - p_{\ell,k})\,\delta_0
\;+\;
p_{\ell,k}\,\mathrm{N}(0, 1),
\qquad
\log \gamma_{\ell,k} \sim (1 - \pi_{\ell,k})\,\delta_0
\;+\;
\pi_{\ell,k}\,\mathrm{N}(0, 1),
\]
where $\delta_0$ denotes a point mass at zero and the mixing probabilities
$p_{\ell,k}$ and $\pi_{\ell,k}$ control the prior inclusion probabilities for
excitation and inhibition, respectively.
The temporal decay parameters governing excitation and inhibition are assigned
independent log-normal priors, $\log \eta_{\ell} \sim \mathrm{N}(0, 1)$ and $\log \phi_{\ell} \sim \mathrm{N}(0, 1)$, reflecting weakly informative prior beliefs about the corresponding time scales.

\subsection{Model Assessment}
\label{sec:assess}

Model adequacy is assessed using the random time change theorem (RTCT; \citealp{Brown1988}), which provides a principled diagnostic for one-dimensional point process models. For a process with conditional intensity $\lambda(t \mid \mathcal{H}_t; \boldsymbol{\theta})$, the associated compensator is defined as $\Lambda(t \mid \mathcal{H}_t; \boldsymbol{\theta}) = \int_0^t \lambda(u \mid \mathcal{H}_u; \boldsymbol{\theta})\,du$. Under correct model specification, transforming the observed event times through the compensator yields inter-event increments that follow an $\mathrm{Exp}(1)$ distribution.
In a Bayesian framework, the compensator is a random quantity, as it depends on the model parameters. Given posterior samples $\boldsymbol{\theta}_b$, $b = 1,\dots,B$, we compute the transformed inter-event increments
\begin{align*}
    d^{*}_{b,i}
    =
    \int_{t_{i-1}}^{t_i}
    \lambda(u \mid \mathcal{H}_u; \boldsymbol{\theta}_b)\,du,
    \qquad i = 1,\dots,n.
\end{align*}
If the model is well specified, the posterior distribution of these transformed
increments should be consistent with that of independent $\mathrm{Exp}(1)$
random variables.

Model adequacy is evaluated using Q--Q plots of the posterior mean estimates of the ordered $\{d^{*}_{(i)}\}$ against the theoretical quantiles of the $\mathrm{Exp}(1)$ distribution, with associated posterior credible bands. Close agreement with the $45^\circ$ reference line indicates adequate model fit, whereas systematic deviations suggest potential model misspecification. Comparisons of these diagnostics across competing models also provide informal model comparison (see \citet{Kang2025whaleHawkes} for further details).

In addition, we consider the widely applicable information criterion \citep[WAIC;][]{watanabe2013widely} to assess out-of-sample predictive performance. 
In the Bayesian framework, WAIC can be readily computed from posterior samples and provides a measure that balances model fit and complexity. Smaller values of WAIC indicate better expected predictive performance.

\section{Simulation Experiments}
\label{sec:sim}

In this section, we conduct simulation experiments to evaluate the performance of
the proposed model and to assess its ability to recover the underlying excitation
and inhibition structure. We first describe the data-generating process, followed by a presentation of the corresponding simulation results.

\subsection{Data Simulation}
\label{sec:sim_data}

Event times from the proposed multivariate Hawkes process with additive
excitation and multiplicative inhibition are generated using a thinning-based
procedure adapted from the algorithm of \citet{Lewis1979}. Let $K$ denote the number of marks and let $\lambda_k(t)$ be the conditional intensity for mark $k$. The simulation proceeds sequentially in continuous time as follows.
\begin{enumerate}

\item \textit{Initialization.}
Set the current time $t = 0$ and initialize the event history
$\mathcal{T} = \emptyset$.

\item \textit{Upper bound construction.}
At time $t$, construct an upper bound $M(t)$ such that
\[
\sum_{k=1}^K \lambda_k(t) \le M(t).
\]
The bound is obtained by combining the maximum background intensity and an upper
bound on the excitation contributions from past events with positive excitation
effects.

\item \textit{Candidate event time.}
Sample a waiting time $w \sim \mathrm{Exp}(M(t))$ and set
$t^\ast = t + w$.
If $t^\ast > T$, terminate the algorithm.

\item \textit{Intensity evaluation.}
For each mark $k = 1, \dots, K$, evaluate the conditional intensity at $t^\ast$, $\lambda_k(t^\ast)$.

\item \textit{Thinning step.}
Accept the candidate event with probability
\[
\frac{\sum_{k=1}^K \lambda_k(t^\ast)}{M(t)}.
\]
If the event is rejected, set $t = t^\ast$ and return to Step~2.

\item \textit{Mark assignment.}
If the event is accepted, sample the event mark $k^\ast$ from a categorical
distribution with probabilities proportional to
$\{\lambda_1(t^\ast), \dots, \lambda_K(t^\ast)\}$.
Add the event $(t^\ast, k^\ast)$ to the history $\mathcal{T}$, set $t = t^\ast$,
and return to Step~2.
\end{enumerate}
The algorithm iterates until the terminal time $T$ is reached.

We generate simulated datasets from three models: (i) the proposed model with both excitation and inhibition, $\lambda_k(t) = \{ \mu_k(t) + G_k(t) \}\, H_k(t)$; (ii) a model with excitation only, $\lambda_k(t) = \mu_k(t) + G_k(t)$; and (iii) a model with inhibition only, $\lambda_k(t) = \mu_k(t)\, H_k(t)$. Analogous to the real data analysis below, we set the number of marks to $K = 3$. The true parameter values are taken to be the posterior median estimates obtained by fitting model (i) to the meerkat dataset in Section \ref{sec:meerkat}. Specifically, the excitation and inhibition matrices are given by
\begin{align*}
    \boldsymbol{\alpha}
    &=
    \begin{pmatrix}
    0.73 & 0.00 & 0.00 \\
    0.00 & 0.90 & 0.26 \\
    0.00 & 0.00 & 0.94
    \end{pmatrix},
    \qquad
    \boldsymbol{\gamma}
    =
    \begin{pmatrix}
    0.00 & 0.00 & 0.26 \\
    0.00 & 0.00 & 0.00 \\
    0.14 & 0.00 & 0.00
    \end{pmatrix}.
\end{align*}
The excitation and inhibition decay parameters are specified as $\boldsymbol{\eta} = (22.38,\; 4.19,\; 2.49)^\top$ and $\boldsymbol{\phi} = (2.64,\; 25.15,\; 2.74)^\top$, respectively.

\begin{figure}[!b]
    \centering
    \includegraphics[width=0.9\linewidth]{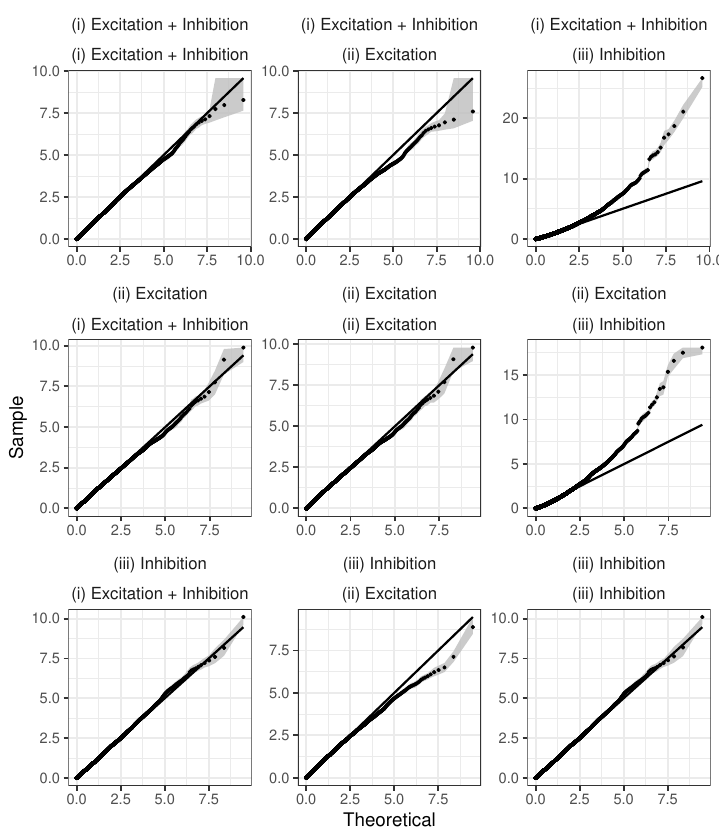}
    \caption{Q-Q plots for $\hat{d}^{\ast}_{(i)}$ against an Exp(1) distribution. Gray shades represent 95\% credible bands. Top and bottom labels denote generating and fitting models, respectively.}
    \label{fig:simQQband}
\end{figure}

\subsection{Results}

We fit models (i) to (iii) to each of the simulated datasets using a Metropolis--Hastings-within-Gibbs sampler for 50{,}000 iterations. The first 10{,}000 iterations were discarded as burn-in, and the remaining 40{,}000 samples were retained for posterior inference. Convergence was assessed through visual inspection of trace plots; no evidence of lack of convergence was observed.

\begin{table}[!b]
\caption{
Posterior median estimate (95\% HPD) of $-2 \log L(\btheta \mid \mathcal{T})$,  WAIC, and mean squared distance (MSD) between sample and theoretical quantiles for models (i) to (iii) fitted to data generated from each model. 
All WAIC values are scaled by $10^4$. 
Boldface indicates the fitting models with relatively smaller MSD or WAIC values.
\label{tab:simMSDnWAIC}}
\begin{center}
\begin{tabular}{llrrrr}
  \toprule
Generating & Fitting & $\widehat{-2 \log L}$ & (95\% HPD) & WAIC & MSD\\ 
  \midrule
  \multirow{3}{*}{(i) Exc + Inh} & (i) Exc + Inh & 23568 & (23557, 23582) & \textbf{2.36} & \textbf{0.001} \\ 
  & (ii) Exc & 23687 & (23679, 23700) & 2.37 & 0.005 \\ 
  & (iii) Inh & 27997 & (27978, 28011) & 2.80 & 0.283 \\ 
  \hline
  \multirow{3}{*}{(ii) Exc} & (i) Exc + Inh & 22527 & (22517, 22538) & \textbf{2.25} & \textbf{0.002} \\ 
  & (ii) Exc & 22527 & (22518, 22538) & \textbf{2.25} & \textbf{0.002} \\ 
  & (iii) Inh & 26767 & (26761, 26777) & 2.68 & 0.195 \\ 
  \hline
  \multirow{3}{*}{(iii) Inh} & (i) Exc + Inh & 26364 & (26351, 26395) & \textbf{2.64} & \textbf{0.001} \\ 
  & (ii) Exc & 26736 & (26724, 26751) & 2.68 & 0.006 \\ 
  & (iii) Inh & 26389 & (26377, 26400) & \textbf{2.64} & \textbf{0.001} \\
   \bottomrule
\end{tabular}
\end{center}
\end{table}

We use RTCT and WAIC to assess how well fitting models can distinguish generating models. 
Figure~\ref{fig:simQQband} presents Q-Q plots of $\hat{d}^{\ast}_{(i)}$ against the $\mathrm{Exp}(1)$ distribution for each combination of generating and fitting models.  Model~(i), which incorporates both excitation and inhibition, provides a good fit across all simulated datasets. In particular, the corresponding Q-Q plots closely follow the reference line, suggesting that model~(i) adequately captures the distributional features of the data generated under the proposed model as well as the simplified models.
Model~(ii), which includes excitation only, exhibits good agreement with the
$\mathrm{Exp}(1)$ reference distribution when fitted to data generated under
model~(ii). When applied to data generated from models~(i) or~(iii), modest deviations from the reference line are observed, indicating some lack of fit in the presence of inhibitory effects.
Model~(iii), which includes inhibition only, aligns well with the reference line
for data generated under model~(iii). However, substantial departures from the $\mathrm{Exp}(1)$ distribution are evident when model~(iii) is fitted to data generated from models~(i) or~(ii), suggesting that this model is not sufficiently flexible to capture excitatory dynamics. 

Table~\ref{tab:simMSDnWAIC} reports the posterior median estimates (95\% HPD intervals) of $-2 \log L(\boldsymbol{\theta} \mid \mathcal{T})$, along with the WAIC and the mean squared distance (MSD) between the sample and theoretical quantiles, for each combination of generating and fitting models. Overall, both WAIC and MSD tend to select the generating model or its nested counterparts, suggesting that these criteria are effective in distinguishing between the true data-generating model and misspecified alternatives.

\section{Applications}
\label{sec:appl}

In this section, we illustrate the proposed methodology through two acoustic communication applications. We first examine interactions among three selected call types in groups of wild meerkats and then analyze within- and between-species calling dynamics in a two-species baleen whale dataset. These applications demonstrate how the additive–multiplicative specification enables interpretable inference on excitation and inhibition in multivariate event sequences.

\subsection{Analysis of Meerkat Call Data}
\label{sec:meerkat}

\begin{figure}[!t]
    \centering
    \includegraphics[width=0.9\linewidth]{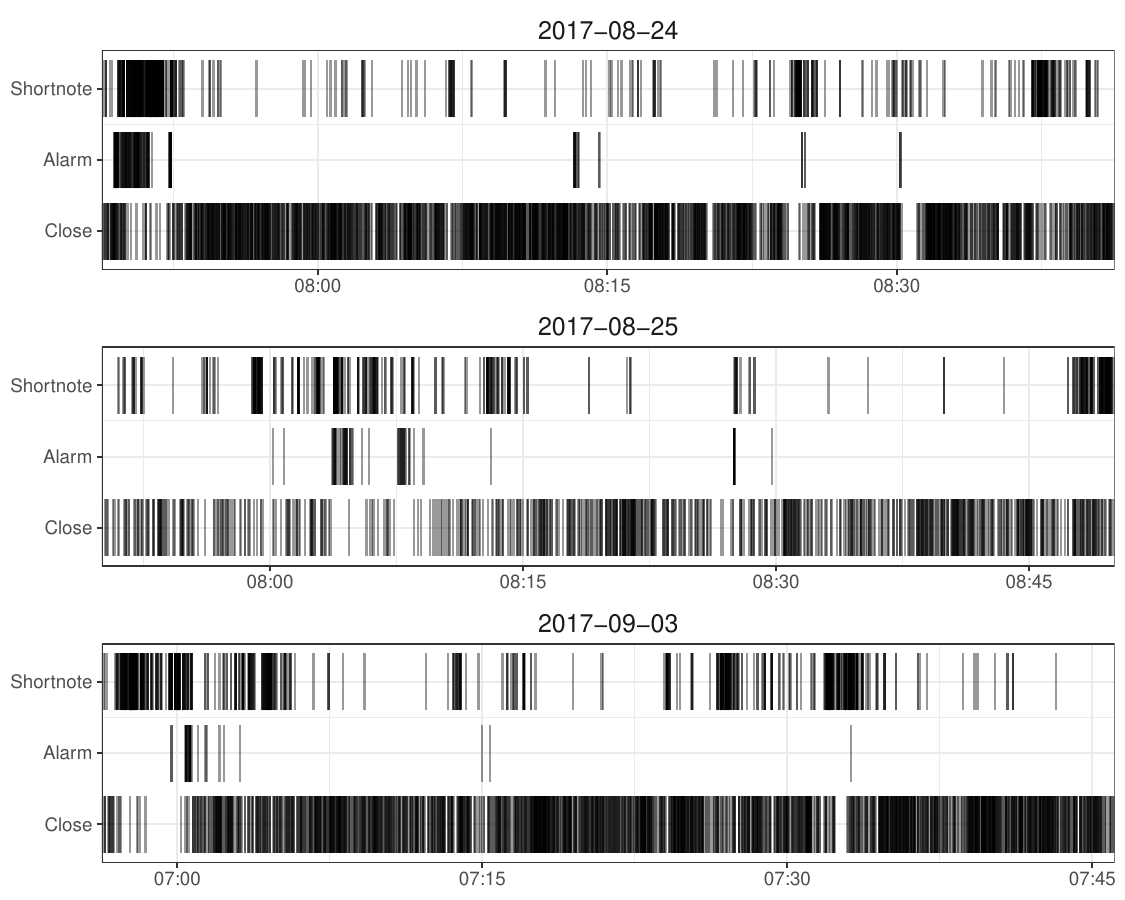}
    \caption{Observed call times by call type (y-axis) over time (x-axis) for three recorded days (facet panels) in 2017 for the meerkat dataset. 
    }
    \label{fig:meerkat_data}
\end{figure}

Meerkats (\textit{Suricata suricatta}) are terrestrial mammals that live in groups in southern Africa, and that have a rich and varied vocal repertoire \citep{manserChapterSixVocal2014}.  Field observations and playback studies over the past several decades have helped uncover the behavioral functions of different call types \citep{manserInformationThatReceivers2001, townsendAllClearMeerkats2011}. More recently, it has become possible to tag multiple individuals simultaneously using tracking collars, enabling continuous vocal sequences to be recorded over several hours from entire social groups \citep{demartsevSignallingGroupsNew2023}, affording opportunities to explore the spatial and temporal dynamics of vocal interactions. Here we use data from a previously published study on meerkats that were tagged and followed while foraging and moving as a group, at the Kalahari Research Centre in South Africa \citep{Demartsev2024}. We focus our analyses on three specific types of meerkat call: (i) the close call, (ii) the alarm call, and (iii) the short note call. Close calls are known to be involved in maintaining group cohesion, and are frequently produced by individuals while foraging \citep{gall2017group,engesser2022collective}. Alarm calls are produced in response to predator detection, and can elicit responses ranging from brief vigilance to running to underground shelters \citep{manser2001acoustic,townsend2012flexible}. Finally, short note calls are used in a variety of contexts, including being produced by raised guards \citep{rauber2017discrete}, while sunning \citep{demartsev2018vocal}, during submission displays towards dominants \citep{reber2013social}, and while running \citep{Demartsev2024}. Although the data were originally collected at the individual level, for analytical tractability we aggregate across all individuals and analyze the event sequences of the three call types across the full group, over three separate days in 2017 (Figure~\ref{fig:meerkat_data}). Table~\ref{tab:meerkat_count} summarizes the observed call counts by type for each day.

To investigate how meerkats respond to different call types, we apply the three models (i)–(iii) described in Section~\ref{sec:sim_data} to the observed event time sequences. Time is measured in seconds.
We assume that the data are observed independently across days. The background intensities are allowed to vary by day and call type, denoted by $\mu_{d,k}$ for days $d = 1,2,3$ and call types $k = 1,2,3$. In contrast, the excitation and inhibition components are assumed to be shared across days, so that the interaction and decay parameters $\alpha_{\ell,k}$, $\gamma_{\ell,k}$, $\eta_{\ell}$, and $\phi_{\ell}$, for $\ell,k = 1,2,3$, taken to be common across all days.

\begin{table}[!t]
\caption{
Observed counts of calls by call type for each of the three recorded days in 2017 for the meerkat dataset.
\label{tab:meerkat_count}}
\begin{center}
\begin{tabular}{lrrr}
  \toprule
 Date & Close & Alarm & Short note \\ 
  \midrule
  2017-08-24 & 1793 & 133 & 550 \\ 
  2017-08-25 & 752 &  63 & 398 \\ 
  2017-09-03 & 1647 &  32 & 700 \\ 
   \bottomrule
\end{tabular}
\end{center}
\end{table}

Posterior inference is carried out using a Metropolis--Hastings-within-Gibbs sampler with 20{,}000 iterations. The first 10{,}000 iterations are discarded as burn-in, and the remaining 10{,}000 samples are retained for inference. Convergence is assessed via visual inspection of trace plots, with no evidence of lack of convergence observed. Likelihood evaluations are parallelized across 20 CPU cores. On the Duke Compute Cluster equipped with Intel Xeon Gold 6252 processors, the model (i) required approximately 5.3 hours to fit, while models (ii) and (iii) required approximately 3.0 and 3.8 hours, respectively.

\begin{figure}[!t]
    \centering
    \includegraphics[width=0.9\linewidth]{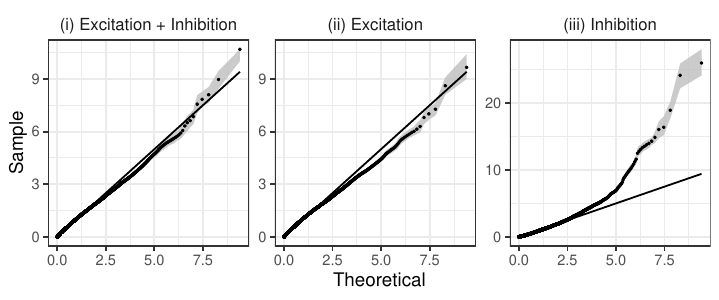}
    \caption{Q-Q plots for $\hat{d}^{\ast}_{(i)}$ against an Exp(1) distribution. Gray shades represent 95\% credible bands. Facet labels denote models fitted to the meerkat dataset.}
    \label{fig:meerkat_rtct}
\end{figure}

Model adequacy for the meerkat dataset is assessed using the randomized time
change theorem (RTCT). Figure~\ref{fig:meerkat_rtct} presents Q-Q plots of samples $\hat{d}^{\ast}_{(i)}$ against the $\mathrm{Exp}(1)$ distribution, with gray shaded regions indicating 95\% credible bands. Model~(i), which incorporates both excitation and inhibition, shows close agreement with the reference line, with only minor deviations observed in the upper tail. Model~(ii), which includes excitation only, also exhibits reasonable alignment with the reference line, although moderate deviations are apparent in the central quantiles. In contrast, model~(iii), which includes inhibition only, displays substantial departures from the reference line across a wide range of quantiles, with the credible band failing to cover the reference line, suggesting a lack of fit for this dataset.


\begin{table}[!b]
\caption{
Posterior median estimate (95\% HPD) of $-2 \log L(\btheta \mid \mathcal{T})$, WAIC, and MSD between sample and theoretical quantiles for models (i) and (ii) fitted to the meerkat dataset. 
All WAIC values are scaled by $10^4$. 
\label{tab:meerkat_modelselection}}
\begin{center}
\begin{tabular}{lrrrr}
  \toprule
Fitting model & $\widehat{-2 \log L}$ & (95\% HPD) & WAIC & MSD\\ 
  \midrule
  (i) Excitation + Inhibition & 20676 & (20663, 20689) & 2.07 & 0.0081 \\ 
  (ii) Excitation & 20773 & (20765, 20785) & 2.08 & 0.0128 \\ 
   \bottomrule
\end{tabular}
\end{center}
\end{table}

Table~\ref{tab:meerkat_modelselection} reports the posterior median estimates
(95\% HPD intervals) of $-2 \log L(\boldsymbol{\theta} \mid \mathcal{T})$, together with the WAIC and MSD values, for models (i) and (ii), both of which appear adequate for the dataset based on the RTCT analysis.
Model~(i) has a significantly smaller value of $-2 \log L(\btheta \mid \mathcal{T})$ than the other models. It also yields comparatively smaller WAIC and MSD values, suggesting improved overall fit relative to the competing model. Based on these criteria, we proceed with model~(i), which incorporates both excitation and inhibition, and present the corresponding results for the meerkat dataset.

\begin{table}[!t]
\caption{Posterior mean estimates and 95\% HPD intervals for interaction parameters ($\alpha_{\ell,k}$ and $\gamma_{\ell,k}$) and decay parameters ($\eta_{\ell}$ and $\phi_{\ell}$) whose HPD intervals exclude zero, for the meerkat dataset.\label{tab:meerkat_parameters}}
\begin{center}
\begin{tabular}{lrr lrr}
  \toprule
Parameter & Mean & 95\% HPD & Parameter & Mean & 95\% HPD \\ 
  \midrule
$\alpha_{\text{cc,cc}}$ & 0.72 & (0.64, 0.80) & $\gamma_{\text{cc,sn}}$ & 0.31 & (0.19, 0.46) \\ 
  $\alpha_{\text{al,al}}$ & 0.90 & (0.78, 1.04) & $\gamma_{\text{sn,cc}}$ & 0.21 & (0.10, 0.32) \\ 
  $\alpha_{\text{al,sn}}$ & 0.25 & (0.14, 0.38) & $\phi_{\text{cc}}$ & 2.18 & (1.10, 3.51) \\ 
  $\alpha_{\text{sn,sn}}$ & 0.93 & (0.87, 1.00) & $\phi_{\text{sn}}$ & 1.84 & (0.77, 3.15) \\ 
  $\eta_{\text{cc}}$ & 21.35 & (16.92, 25.48) &  &  &  \\ 
  $\eta_{\text{al}}$ & 4.06 & (2.97, 5.18) &  &  &  \\ 
  $\eta_{\text{sn}}$ & 2.57 & (2.14, 3.05) &  &  &  \\ 
   \bottomrule
\end{tabular}
\end{center}
\end{table}

Table~\ref{tab:meerkat_parameters} reports the posterior mean estimates and 95\% HPD intervals for interaction and decay parameters whose HPD intervals exclude zero, providing evidence for the presence of corresponding excitatory or inhibitory effects. 
The magnitudes of the excitation parameters $\alpha_{\ell,k}$ indicate that self-excitation is strongest for alarm and short note calls, whereas the corresponding effect for close calls is more moderate, with posterior means of $\alpha_{\mathrm{al,al}} = 0.90$, $\alpha_{\mathrm{sn,sn}} = 0.93$, and $\alpha_{\mathrm{cc,cc}} = 0.72$. In addition, cross-type excitation from alarm to short note is relatively small in magnitude, with a posterior mean of $\alpha_{\mathrm{al,sn}} = 0.25$. 
The excitation decay parameters $\eta_{\ell}$ characterize the average temporal duration over which an excitation effect from call type $\ell$ persists. These durations differ significantly across call types. In particular, excitation effects associated with close calls persist for approximately 21 seconds on average, indicating relatively long-lasting influence. In contrast, excitation effects from alarm calls decay more rapidly, with an average duration of about 4 seconds, while those from short note calls are even shorter, lasting approximately 3 seconds on average. 
The inhibition parameters $\gamma_{\ell,k}$ quantify the magnitude of inhibition effects between call types. The posterior estimates provide evidence of inhibitory interactions in both directions between close and short note calls, with posterior means of $\gamma_{\mathrm{cc,sn}} = 0.31$ and $\gamma_{\mathrm{sn,cc}} = 0.21$. However, the substantial overlap of the corresponding 95\% HPD intervals precludes a definitive comparison of inhibition strength across the two directions. 
The inhibition decay parameters $\phi_{\ell}$ characterize the temporal range over which inhibitory effects operate. Although these parameters do not admit a direct interpretation analogous to the excitation decay parameters $\eta_{\ell}$, the estimated values for close and short note calls, $\phi_{\mathrm{cc}} = 2.18$ and $\phi_{\mathrm{sn}} = 1.84$, indicate that inhibitory effects associated with these call types act over similar temporal ranges.

\begin{figure}[!t]
    \centering
    \begin{tikzpicture}[
        >=Stealth,
        every node/.style = {font=\small},
        call/.style   = {circle, draw, thick, minimum size=2cm},
        alarm/.style = {call, fill=pastelred},
        short/.style = {call, fill=pastelblue},
        close/.style = {call, fill=pastelgreen},
        excite/.style  = {->},
        inhibit/.style = {->, dashed}
    ]
    
    \node[alarm] (al)  {Alarm};
    \node[short, right=4.4cm of al] (sn) {short note};
    \node[close, right=1cm of al, yshift=-3.5cm] (cc) {Close};

    \draw[excite, line width={exp(0.90*1.2)}] (al) edge[loop left, looseness=4] (al);
    \node[left=6mm of al] {\makecell{Self-excitation \vspace{-0.1in} \\ (0.90)}};
    
    \draw[excite, line width={exp(0.93*1.2)}] (sn) edge[loop right, looseness=4] (sn);
    \node[right=6mm of sn] {\makecell{Self-excitation \vspace{-0.1in} \\ (0.93)}};
    
    \draw[excite, line width={exp(0.72*1.2)}] (cc) edge[loop below, looseness=4] (cc);
    \node[below=6mm of cc] {Self-excitation (0.72)};

    \draw[excite, line width={exp(0.25*1.2)}] (al) -- node[above=2pt] {Cross-excitation (0.25)}
                      (sn);
    
    \draw[inhibit, bend right=10, line width={exp(0.21*1.2)}] (sn) to 
    node[pos=0.3, left=0.1pt] {\makecell{Cross-inhibition \vspace{-0.1in} \\ (0.21)}}
                         (cc);
    
    \draw[inhibit, bend right=10, line width={exp(0.31*1.2)}] (cc) to 
    node[pos=0.3, right=10pt] {\makecell{Cross-inhibition \vspace{-0.1in} \\ (0.31)}}
    (sn);
    
    \end{tikzpicture}
    \caption{Excitatory (solid arrows) and inhibitory (dashed arrows) interactions among alarm, short note, and close calls in the meerkat dataset.}
    \label{fig:meerkat_diagram}
\end{figure}
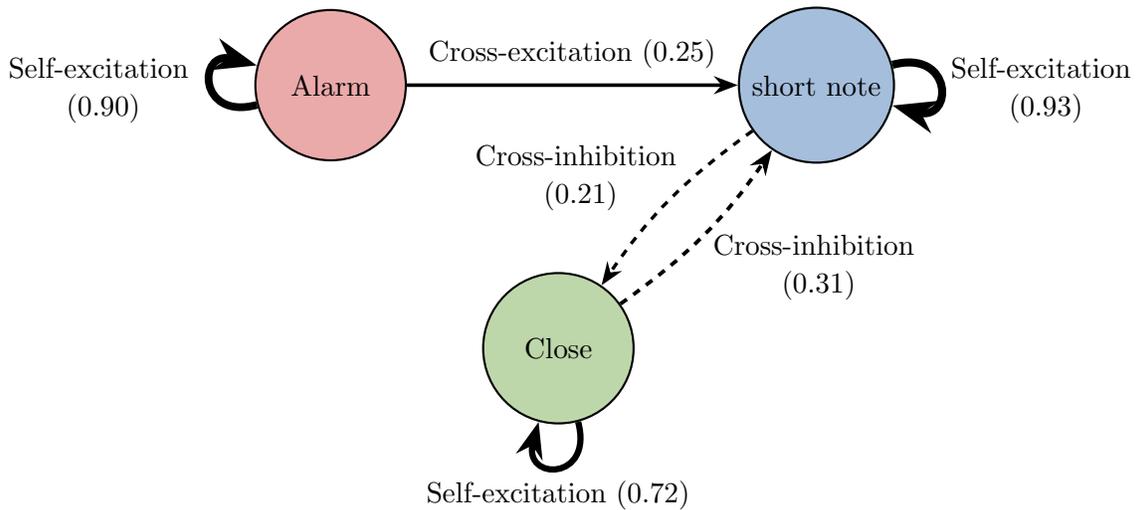

Figure~\ref{fig:meerkat_diagram} depicts the inferred interaction structure among alarm, short note, and close calls in the meerkat dataset. This indicates that alarm calls tend to be followed by additional alarm calls through self-excitation and also promote the occurrence of short note calls. Short note calls likewise show self-excitation, but are associated with a reduced occurrence of close calls, reflecting an inhibitory effect. In contrast, close calls are self-exciting and decrease the occurrence of short note calls, suggesting that close and short note calls tend to suppress each other’s occurrence. The fitted parameters for excitation and inhibition are consistent with biological expectations. Close calls, which are produced in a foraging context, have already been shown via both observations and playback experiments to be self-excitatory in that they induce increased calling in nearby conspecifics \citep{engesser2022collective}. Short note and alarm calls are also often produced in sequences. Because short note calls are typically produced in contexts other than foraging \citep{rauber2017discrete,demartsev2018vocal,reber2013social,Demartsev2024}, the inhibition effect between short notes and close calls captured by our model could reflect that these calls are used in different behavioral contexts that are mutually exclusive with another (e.g., foraging vs running, foraging vs raised guarding behavior). Finally, the cross-excitation between alarm calls and short note recalls could reflect sequences often observed in meerkats, where alarm calls are produced followed by running to shelter (while producing short note calls) \citep{Demartsev2024}. It is important to note that this model ignores important biological considerations such as which individuals are calling and their spatial configuration. Despite these limitations, the finding that even this relatively simplistic model is able to capture biologically meaningful interactions among call types in a well-understood system such as meerkats highlights the potential of our modeling approach for gaining useful insights into animal communication.

\subsection{Analysis of Whale Call Data}
\label{sec:whale}
Right whales are known to make a stereotyped frequency-modulated upcall \citep{clarkAcousticRepertoireSouthern1982} that serves as a contact call \citep{clarkSoundPlaybackExperiments1980}. Using passive acoustic monitoring (PAM), this has been successfully used to document right whale presence and absence across their habitat range \citep{davisUpcallingBehaviourPatterns2023}, and more recently has been used to examine calling and counter-calling behavior across a foraging habitat \citep{Kang2025whaleHawkes}. Humpback whales are known to display a range of sophisticated acoustic behavior \citep{paynemcvay1971}.

\begin{figure}[!b]
    \centering
    \includegraphics[width=0.9\linewidth]{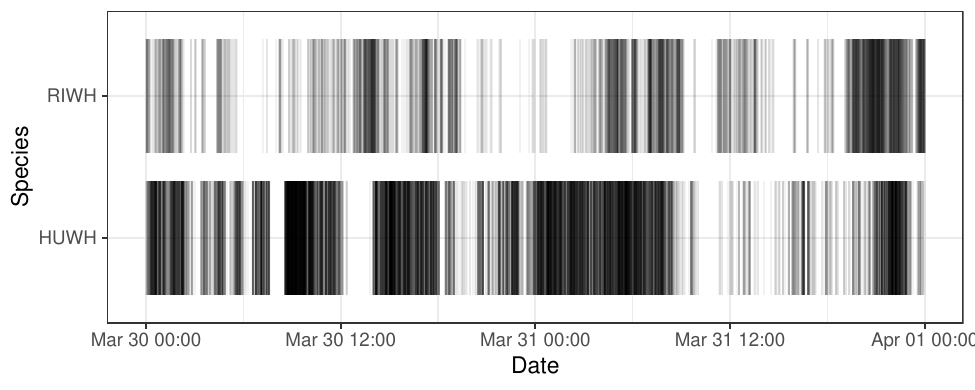}
    \caption{Observed call times by species (y-axis) over the two-day period from March 30 to March 31, 2009. Each vertical line represents the occurrence time of an individual call, with time shown on the x-axis. 
    }
    \label{fig:whale_calltimes}
\end{figure}

Because PAM is such an active area of research, a biennial conference exists to discuss research and algorithmic development---the Detection, Classification, Localization, and Density Estimation of marine mammals conference (DCL/DE). At each conference, a challenge dataset is made available for researchers to use and compare algorithmic development. Here we use the weeklong dataset from the 2013 conference, which initially focused on right whale upcalls detected from a buoy in Stellwagen Bank National Marine Sanctuary, MA. The dataset was later expanded to include other species of baleen whales \citep{dclde2013}, using the Low Frequency Detection and Classification System from \citet{baumgartnerGeneralizedBaleenWhale2011}. We limit our analysis to the two-day period from March 30 to March 31, 2009, when a total of 6,068 calls for humpback whales (HUWH), and 2,492 calls for North Atlanic right whales (RIWH) were recorded (Figure~\ref{fig:whale_calltimes}). LFDCS includes 6 types of frequency modulated HUWH calls \citep{baumgartnerRealtimeReportingBaleen2013}, as well as the frequency modulated RIWH upcall. In addition to the animal recordings, the PAM deployments also measure ambient noise, which allows us to explore the role noise may play in masking or inhibiting calling behavior \citep{matthewsOverviewNorthAtlantic2021}.


\begin{figure}[!t]
    \centering
    \includegraphics[width=0.9\linewidth]{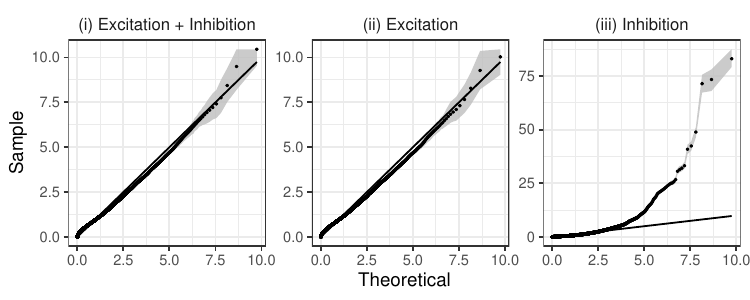}
    \caption{Q-Q plots for $\hat{d}^{\ast}_{(i)}$ against an Exp(1) distribution. Gray shades represent 95\% credible bands. Facet labels denote models fitted to the whale dataset.}
    \label{fig:whale_rtct}
\end{figure}

To investigate how whales respond to calls from conspecifics and heterospecifics, we apply models (i)--(iii) described in Section~\ref{sec:sim_data} to the observed call time sequences. Time is measured in minutes. The background process $\mu_k(t; \boldsymbol{\theta}_{\mu})$ represents the baseline rate of contact calls for species $k$ at time $t$. 
To account for the effect of ambient noise on calling behavior, as well as residual temporal inhomogeneity in contact calls, we specify
\begin{align}
    \log \mu_k(t; \boldsymbol{\theta}_{\mu})
    =
    \beta_{0k}
    +
    \beta_{\text{noise},k}\,\text{Noise}(t)
    +
    W_k(t),
    \nonumber
\end{align}
where $\text{Noise}(t)$ denotes the ambient noise level at time $t$. 
The term $W_k(t)$ is modeled as a mean-zero Gaussian process with an exponential covariance function, characterized by variance parameter $\kappa$ and range parameter $\rho$, allowing for smooth temporal deviations from the parametric mean structure.
In our implementation, the range parameter $\rho$ of the Gaussian process is fixed at two hours. This specification is intended to capture smooth, broad-scale temporal variation in the background intensity, rather than short-term local fluctuations. Constraining $\rho$ at this scale also reduces potential confounding between the Gaussian process component and the excitation and inhibition effects, which are designed to model comparatively shorter-term interaction dynamics.
The excitation component characterizes countercalling behavior, while the inhibition component captures suppressive effects of recent calls that reduce subsequent calling activity.


\begin{table}[!t]
\caption{
Posterior median estimate (95\% HPD) of $-2 \log L(\btheta \mid \mathcal{T})$, WAIC, and mean squared distance (MSD) between sample and theoretical quantiles for models (i) and (ii) fitted to the whale dataset. 
All WAIC values are scaled by $10^4$. 
\label{tab:whale_modelselection}}
\begin{center}
\begin{tabular}{lrrrr}
  \toprule
Fitting model & $\widehat{-2 \log L}$ & (95\% HPD) & WAIC & MSD\\ 
  \midrule
  (i) Excitation + Inhibition &   -99 & ( -147,   -2) &   0.01 & 0.0141 \\ 
  (ii) Excitation &   -66 & ( -118,   -3) &   0.01 & 0.0143 \\ 
   \bottomrule
\end{tabular}
\end{center}
\end{table}

Posterior inference is performed using a Metropolis--Hastings within-Gibbs sampler with 20{,}000 iterations, of which the first 10{,}000 are discarded as burn-in. The remaining 10{,}000 samples are used for inference. Convergence is assessed through trace plots, with no apparent issues detected. Likelihood evaluations are parallelized over 20 CPU cores. On the Duke Compute Cluster with Intel Xeon Gold 6252 processors, model~(i) required approximately 20 hours of computation, whereas models~(ii) and~(iii) required about 15 and 18 hours, respectively.

Model adequacy for the whale dataset is assessed using the RTCT, with results shown in
Figure~\ref{fig:whale_rtct}. Models~(i) and~(ii) closely follow the reference
line, whereas model~(iii) exhibits substantial deviations, indicating a lack of
fit for this dataset.
Table~\ref{tab:whale_modelselection} presents posterior median estimates (95\% HPD intervals) of $-2 \log L(\boldsymbol{\theta} \mid \mathcal{T})$ together with WAIC and MSD values, for models (i) and (ii), both of which appear adequate for the dataset based on the RTCT analysis. 
Models~(i) and~(ii) yield similar values for these criteria, which indicates that models~(i) and~(ii) provide adequate fits to the whale dataset. 
Given their comparable performance, we select the more parsimonious specification, model~(ii), which includes excitation only, and present the corresponding results below.

\begin{table}[!t]
\caption{Posterior mean estimates and 95\% HPD intervals for the noise coefficients ($\beta_{\text{noise},k}$), interaction parameters ($\alpha_{\ell,k}$), and decay parameters ($\eta_{\ell}$) whose HPD intervals exclude zero, for the whale dataset.\label{tab:whale_parameters}}
\begin{center}
\begin{tabular}{lrr lrr}
  \toprule
Parameter & Mean & 95\% HPD & Parameter & Mean & 95\% HPD \\ 
  \midrule
$\beta_{\text{\tiny{noise,RIWH}}}$ & -0.09 & (-0.30, 0.11) & $\alpha_{\text{\tiny{RIWH,RIWH}}}$ & 0.80 & (0.72, 0.89) \\ 
  $\beta_{\text{\tiny{noise,HUWH}}}$ & -0.26 & (-0.37, -0.14) & $\alpha_{\text{\tiny{HUWH,HUWH}}}$ & 0.62 & (0.58, 0.66) \\ 
  $\eta_{\text{\tiny{RIWH}}}$ & 2.86 & (2.34, 3.38) &  &  &  \\ 
  $\eta_{\text{\tiny{HUWH}}}$ & 0.47 & (0.43, 0.52) &  &  &  \\ 
   \bottomrule
\end{tabular}
\end{center}
\end{table}

Table~\ref{tab:whale_parameters} reports posterior mean estimates and 95\% HPD intervals for the noise coefficients, as well as interaction and decay parameters whose HPD intervals exclude zero, providing evidence for the presence of corresponding excitatory effects.
The posterior estimates of the noise coefficients are negative for both species. For humpback whales, the 95\% HPD interval lies entirely below zero, indicating a statistically significant negative effect of ambient noise on calling intensity.

The estimated excitation parameters $\alpha_{\ell,k}$ suggest strong
self-excitation for both species, with posterior means of $\alpha_{\text{\tiny{RIWH,RIWH}}} = 0.80$ and $\alpha_{\text{\tiny{HUWH,HUWH}}} = 0.62$.
The excitation decay parameters $\eta_{\ell}$ describe how long, on average, the excitatory influence of a call from species $\ell$ persists over time. The estimated values demonstrate pronounced differences between the two species. Excitation induced by right whale calls persists for an average of 2.86 minutes, while excitation associated with humpback whale calls decays considerably more quickly, with an average duration of approximately 28 seconds. This estimate is slightly shorter than that reported in \citep{Kang2025whaleHawkes}, which examined RIWH upcalls in Cape Cod Bay (CCB), MA. As compared to CCB, whales near the location in this analysis are more likely to be transiting then aggregating. Thus it stands to reason the decay is shorter here. The single RIWH call type analyzed here is a known contact call \citep{clarkAcousticRepertoireSouthern1982}. In contrast, the call data analyzed for HUWH comprise 6 different call types, including song that is only produced by males and typically serves a breeding function. It follows, therefore, that excitation effect might be lower in HUWH. Though the breeding-related songs are very well known and characterized \citep{paynemcvay1971}, HUWH are also known to produce a diverse array of sounds on foraging grounds \citep{dunlop2022humpback}. Some of these sounds are presumed to either alter the behavior of prey, or to communicate to nearby conspecifics \citep{parksEvidenceAcousticCommunication2014}. Combined with the diversity of HUWH call types analyzed here, each of which may have a different behavioral role, this potential dual role of foraging related sounds may account for the slightly lower excitement estimate for HUWH. Because social foraging has been postulated in RIWH \citep{sorochanAvailabilitySupplyAggregation2021} and observed in HUWH \citep{findlay2017humpback}, it would be interesting to couple these acoustic records with longer visual observations to better understand the role of these call types within and across individuals.

\section{Summary and Future Work}

We have studied the modeling of excitation and inhibition in animal acoustic communication. While Hawkes processes provide a natural framework for representing excitation, their application in animal communication remains relatively limited. Incorporating inhibition within this framework is nontrivial, as the conditional intensity must remain nonnegative, and identifiability challenges can arise when modeling unmarked event sequences.

We have offered a multivariate model featuring within and across component additive excitement with multiplicative inhibition. 
This specification offers numerous attractive features including a clear separation of processes, direct interpretation of parameters, avoidance of link function specification, improved parameter identifiability, and efficient MCMC model fitting.  We have illuminated the performance of this class of models through both simulation and two real data examples.

The proposed model and analysis has captured and quantified ecologically meaningful structure in the multivariate event sequences of three different mammalian species. 
Our approach has been able to quantify how this call type diffuses through the system, and our results are consistent with empirical observations of both excitatory and inhibitory acoustic behavior \citep{Demartsev2024}. In contrast, no significant inhibitory communication effects were identified in the heterospecific whale data. Though social foraging is presumed to exist in baleen whales \citep{cadeSocialExploitationExtensive2021}, and eavesdropping behavior has been documented in humpback whales \citep{dunlopEavesdroppingRiskConspicuous2021}, the foraging niches of humpback and right whales do not overlap \citep{goldbogenIntegrativeApproachesStudy2013}. Therefore, it is not surprising that a presumed socially-mediated foraging aggregation of one species does not lead to corresponding changes in the acoustic behavior of the other \citep{seppanenSocialInformationUse2007}. While we do not know if the negative coefficients for noise for each species are owing to a masking effect \citep{clarkAcousticMaskingMarine2009} or an inhibitory effect, or both \citep{urazghildiievStatisticalAnalysisNorth2014}, acoustic communication is critical for all marine mammal species \citep{tyackCommunicationAcousticBehavior2000}. Thus, any negative impact of increased ambient noise in an industrializing ocean warrants both further research and conservation action \citep{duarteSoundscapeAnthropoceneOcean2021a}.

Important future work which we are currently exploring is the introduction of spatial dependence into our modeling.  Hawkes excitement processes have been expanded to allow for excitement to arise from events recent in time and in close spatial proximity. Customarily, the trigger function is specified as the product of a temporal excitement term and a spatial excitement term \citep{Reinhart2018}.  We are investigating how to incorporate such behavior into our ``additive by multiplicative'' specification for excitement and inhibition.  With multivariate process modeling, evident challenges across processes include disentangling temporal and spatial behavior as well as identifying conflict between, e.g., temporal excitement and spatial inhibition or vice versa.

On-going exploration with the meerkat data looks to further partition the call type data at the individual level. This would enable more resolved inference pertaining to within- and cross-individual excitation within call type. 
Revision of our modeling would introduce a substantial number of additional parameters and identifiability constraints and is beyond our scope of the work presented here. Similarly, while the acoustic records for the RIWHs and HUWDs included only one call type, they are known to make many different types of behaviorally specific calls \citep{clarkAcousticRepertoireSouthern1982,matthewsOverviewNorthAtlantic2021}. Expanding the dataset to include additional call types and extending the model could yield better understanding of the call-to-call influence, and possibly help explain the interspecific differences in the decay parameters that we observed. 
Within the field of marine mammal bioacoustics, the study of the acoustic response of prey to killer whale predators would also be a rich area to explore from an ecological \citep{branchMostFlightBaleen2025}, evolutionary \citep{morisakaPredationKillerWhales2007}, and conservation perspective \citep{millerBehavioralResponsesPredatory2022}.

Apart from crime data, as considered in \citet{Olinde2020}, the proposed class of models may be applicable in various other domains. Potential areas for future investigation include finance (e.g., market events and trading activity), epidemiology (e.g., disease transmission and contagion processes), and social networks (e.g., information diffusion and opinion dynamics).

\section*{Acknowledgements}
We thank Gen Davis and the Passive Acoustic Branch at NOAA Fisheries' Northeast Fisheries Science Center for sharing the marine mammal acoustic data. We acknowledge support from the US Office of Naval Research (Awards: N000142312562 and N000142412501).



\bibliographystyle{apalike}
\bibliography{refs}

@article{reber2013social,
  title={{Social monitoring via close calls in meerkats}},
  author={Reber, Stephan A and Townsend, Simon W and Manser, Marta B},
  journal={Proceedings of the Royal Society B: Biological Sciences},
  volume={280},
  number={1765},
  year={2013},
  publisher={The Royal Society}
}

@article{demartsev2018vocal,
  title={{Vocal turn-taking in meerkat group calling sessions}},
  author={Demartsev, Vlad and Strandburg-Peshkin, Ariana and Ruffner, Michaela and Manser, Marta},
  journal={Current Biology},
  volume={28},
  number={22},
  pages={3661--3666},
  year={2018},
  publisher={Elsevier}
}

@article{rauber2017discrete,
  title={{Discrete call types referring to predation risk enhance the efficiency of the meerkat sentinel system}},
  author={Rauber, Ramona and Manser, Marta B},
  journal={Scientific reports},
  volume={7},
  number={1},
  pages={44436},
  year={2017},
  publisher={Nature Publishing Group UK London}
}

@article{manser2001acoustic,
  title={{The acoustic structure of suricates' alarm calls varies with predator type and the level of response urgency}},
  author={Manser, Marta B},
  journal={Proceedings of the Royal Society of London. Series B: Biological Sciences},
  volume={268},
  number={1483},
  pages={2315--2324},
  year={2001},
  publisher={The Royal Society}
}

@article{townsend2012flexible,
  title={{Flexible alarm calling in meerkats: the role of the social environment and predation urgency}},
  author={Townsend, Simon W and Rasmussen, Maria and Clutton-Brock, Tim and Manser, Marta B},
  journal={Behavioral Ecology},
  volume={23},
  number={6},
  pages={1360--1364},
  year={2012},
  publisher={Oxford University Press UK}
}

@article{gall2017group,
  title={{Group cohesion in foraging meerkats: follow the moving `vocal hot spot'}},
  author={Gall, Gabriella EC and Manser, Marta B},
  journal={Royal Society open science},
  volume={4},
  number={4},
  year={2017},
  publisher={The Royal Society}
}

@article{miller2009antiphonal,
  title={{Antiphonal call timing in marmosets is behaviorally significant: interactive playback experiments}},
  author={Miller, Cory T and Beck, Kaylin and Meade, Brooke and Wang, Xiaoqin},
  journal={Journal of Comparative Physiology A},
  volume={195},
  number={8},
  pages={783--789},
  year={2009},
  publisher={Springer}
}

@article{engesser2022collective,
  title={{Collective close calling mediates group cohesion in foraging meerkats via spatially determined differences in call rates}},
  author={Engesser, Sabrina and Manser, Marta B},
  journal={Animal Behaviour},
  volume={185},
  pages={73--82},
  year={2022},
  publisher={Elsevier}
}

@article{Reinhart2018,
   author = {Alex Reinhart},
   doi = {10.1214/17-STS629},
   issn = {08834237},
   issue = {3},
   journal = {Statistical Science},
   month = {8},
   pages = {299-318},
   publisher = {Institute of Mathematical Statistics},
   title = {{A review of self-exciting spatio-temporal point processes and their applications}},
   volume = {33},
   year = {2018}
}

@book{bradburyPrinciplesAnimalCommunication1998,
  title = {Principles of Animal Communication},
  author = {Bradbury, Jack W and Vehrencamp, Sandra L},
  year = 1998,
  publisher = {Sinauer}
}

@article{kershenbaumAcousticSequencesNonhuman2016,
  title = {Acoustic Sequences in Non-Human Animals: A Tutorial Review and Prospectus},
  shorttitle = {Acoustic Sequences in Non-Human Animals},
  author = {Kershenbaum, Arik and Blumstein, Daniel T. and Roch, Marie A. and Ak{\c c}ay, {\c C}a{\u g}lar and Backus, Gregory and Bee, Mark A. and Bohn, Kirsten and Cao, Yan and Carter, Gerald and C{\"a}sar, Cristiane and Coen, Michael and DeRuiter, Stacy L. and Doyle, Laurance and Edelman, Shimon and {Ferrer-i-Cancho}, Ramon and Freeberg, Todd M. and Garland, Ellen C. and Gustison, Morgan and Harley, Heidi E. and Huetz, Chlo{\'e} and Hughes, Melissa and Hyland Bruno, Julia and Ilany, Amiyaal and Jin, Dezhe Z. and Johnson, Michael and Ju, Chenghui and Karnowski, Jeremy and Lohr, Bernard and Manser, Marta B. and McCowan, Brenda and Mercado III, Eduardo and Narins, Peter M. and Piel, Alex and Rice, Megan and Salmi, Roberta and Sasahara, Kazutoshi and Sayigh, Laela and Shiu, Yu and Taylor, Charles and Vallejo, Edgar E. and Waller, Sara and {Zamora-Gutierrez}, Veronica},
  year = 2016,
  journal = {Biological Reviews},
  volume = {91},
  number = {1},
  pages = {13--52},
  issn = {1469-185X},
  doi = {10.1111/brv.12160},
  urldate = {2025-01-28},
  copyright = {\copyright{} 2014 Cambridge Philosophical Society},
  langid = {english}
}

@article{demartsevSignallingGroupsNew2023,
  title = {Signalling in Groups: {{New}} Tools for the Integration of Animal Communication and Collective Movement},
  shorttitle = {Signalling in Groups},
  author = {Demartsev, Vlad and Gersick, Andrew S. and Jensen, Frants H. and Thomas, Mara and Roch, Marie A. and Manser, Marta B. and Strandburg-Peshkin, Ariana},
  year = 2023,
  month = aug,
  journal = {Methods in Ecology and Evolution},
  volume = {14},
  number = {8},
  pages = {1852--1863},
  issn = {2041-210X, 2041-210X},
  doi = {10.1111/2041-210X.13939},
  urldate = {2023-12-01}
}

@article{clarkSoundPlaybackExperiments1980,
  title = {Sound {{Playback Experiments}} with {{Southern Right Whales}} ( {{{\emph{Eubalaena}}}}{\emph{ Australis}} )},
  author = {Clark, Christopher W. and Clark, Jane M.},
  year = 1980,
  month = feb,
  journal = {Science},
  volume = {207},
  number = {4431},
  pages = {663--665},
  issn = {0036-8075, 1095-9203},
  doi = {10.1126/science.207.4431.663},
  urldate = {2024-01-17},
  langid = {english}
}

@article{urazghildiievStatisticalAnalysisNorth2014,
  title = {Statistical Analysis of {{North Atlantic}} Right Whale ({{Eubalaena}} Glacialis) Signal Trains in {{Cape Cod Bay}}, Spring 2012},
  author = {Urazghildiiev, Ildar R.},
  year = 2014,
  month = nov,
  journal = {The Journal of the Acoustical Society of America},
  volume = {136},
  number = {5},
  pages = {2851--2860},
  issn = {0001-4966},
  doi = {10.1121/1.4898048},
  urldate = {2024-02-16}
}

@article{matthewsOverviewNorthAtlantic2021,
  title = {An Overview of {{North Atlantic}} Right Whale Acoustic Behavior, Hearing Capabilities, and Responses to Sound},
  author = {Matthews, Leanna P. and Parks, Susan E.},
  year = 2021,
  month = dec,
  journal = {Marine Pollution Bulletin},
  volume = {173},
  pages = {113043},
  issn = {0025-326X},
  doi = {10.1016/j.marpolbul.2021.113043},
  urldate = {2022-10-28},
  langid = {english}
}

@article{baumgartnerGeneralizedBaleenWhale2011,
  title = {A Generalized Baleen Whale Call Detection and Classification System},
  author = {Baumgartner, Mark F and Mussoline, Sarah E},
  year = 2011,
  month = may,
  journal = {J. Acoust. Soc. Am.},
  volume = {129},
  number = {5},
  pages = {2889--2902},
  issn = {0001-4966},
  doi = {10.1121/1.3562166}
}

@article{davisUpcallingBehaviourPatterns2023,
  title = {Upcalling Behaviour and Patterns in {{North Atlantic}} Right Whales, Implications for Monitoring Protocols during Wind Energy Development},
  author = {Davis, G E and Tennant, S C and Van~Parijs, S M},
  year = 2023,
  month = nov,
  journal = {ICES Journal of Marine Science},
  pages = {fsad174},
  issn = {1054-3139},
  doi = {10.1093/icesjms/fsad174},
  urldate = {2023-11-15}
}

@article{clarkAcousticRepertoireSouthern1982,
  title = {The Acoustic Repertoire of the {{Southern}} Right Whale, a Quantitative Analysis},
  author = {Clark, Christopher W},
  year = 1982,
  journal = {Anim. Behav.},
  volume = {30},
  number = {4},
  pages = {1060--1071},
  issn = {0003-3472},
  doi = {10.1016/s0003-3472(82)80196-6}
}

@incollection{dunlop2022humpback,
  title={Humpback whales: A seemingly socially simple whale with communicative complexity},
  author={Dunlop, Rebecca A},
  editor={C.W. Clark and E. C. Garland},
  booktitle={Ethology and behavioral ecology of mysticetes},
  pages={223--246},
  year={2022},
  publisher={Springer}
}

@article{findlay2017humpback,
  title={Humpback whale ``super-groups''--A novel low-latitude feeding behaviour of Southern Hemisphere humpback whales (Megaptera novaeangliae) in the Benguela Upwelling System},
  author={Findlay, Ken P and Seakamela, S Mduduzi and Me{\"y}er, Michael A and Kirkman, Stephen P and Barendse, Jaco and Cade, David E and Hurwitz, David and Kennedy, Amy S and Kotze, Pieter GH and McCue, Steven A and others},
  journal={PloS one},
  volume={12},
  number={3},
  pages={e0172002},
  year={2017},
  publisher={Public Library of Science}
}

@article{sorochanAvailabilitySupplyAggregation2021,
  title = {Availability, Supply, and Aggregation of Prey ({{Calanus}} Spp.) in Foraging Areas of the {{North Atlantic}} Right Whale ({{Eubalaena}} Glacialis)},
  author = {Sorochan, K A and Plourde, S and Baumgartner, M F and Johnson, C L},
  year = 2021,
  month = dec,
  journal = {ICES Journal of Marine Science},
  volume = {78},
  number = {10},
  pages = {3498--3520},
  issn = {1054-3139},
  doi = {10.1093/icesjms/fsab200},
  urldate = {2022-01-11}
}

@article{parksEvidenceAcousticCommunication2014,
  title = {Evidence for Acoustic Communication among Bottom Foraging Humpback Whales},
  author = {Parks, Susan E. and Cusano, Dana A. and Stimpert, Alison K. and Weinrich, Mason T. and Friedlaender, Ari S. and Wiley, David N.},
  year = 2014,
  month = dec,
  journal = {Scientific Reports},
  volume = {4},
  number = {1},
  pages = {7508},
  publisher = {Nature Publishing Group},
  issn = {2045-2322},
  doi = {10.1038/srep07508},
  urldate = {2026-03-02},
  copyright = {2014 The Author(s)},
  langid = {english}
}

@article{baumgartnerRealtimeReportingBaleen2013,
  title = {Real-Time Reporting of Baleen Whale Passive Acoustic Detections from Ocean Gliders},
  author = {Baumgartner, Mark F. and Fratantoni, David M. and Hurst, Thomas P. and Brown, Moira W. and Cole, Tim V. N. and Van Parijs, Sofie M. and Johnson, Mark},
  year = 2013,
  month = sep,
  journal = {The Journal of the Acoustical Society of America},
  volume = {134},
  number = {3},
  pages = {1814--1823},
  issn = {0001-4966},
  doi = {10.1121/1.4816406},
  urldate = {2026-02-14}
}

@article{cadeSocialExploitationExtensive2021,
  title = {Social Exploitation of Extensive, Ephemeral, Environmentally Controlled Prey Patches by Supergroups of Rorqual Whales},
  author = {Cade, David E. and Fahlbusch, James A. and Oestreich, William K. and Ryan, John and Calambokidis, John and Findlay, Ken P. and Friedlaender, Ari S. and Hazen, Elliott L. and Seakamela, S. Mduduzi and Goldbogen, Jeremy A.},
  year = 2021,
  month = dec,
  journal = {Animal Behaviour},
  volume = {182},
  pages = {251--266},
  issn = {0003-3472},
  doi = {10.1016/j.anbehav.2021.09.013},
  urldate = {2023-09-17}
}

@article{clarkAcousticMaskingMarine2009,
  title = {Acoustic Masking in Marine Ecosystems: Intuitions, Analysis, and Implication},
  author = {Clark, C W and Ellison, W T and Southall, B L and Hatch, L and Van Parijs, S M and Frankel, A and Ponirakis, D},
  year = 2009,
  month = dec,
  journal = {Mar. Ecol. Prog. Ser.},
  volume = {395},
  pages = {201--222},
  issn = {0171-8630},
  doi = {10.3354/meps08402}
}

@article{millerBehavioralResponsesPredatory2022,
  title = {Behavioral Responses to Predatory Sounds Predict Sensitivity of Cetaceans to Anthropogenic Noise within a Soundscape of Fear},
  author = {Miller, Patrick J. O. and Isojunno, Saana and Siegal, Eilidh and Lam, Frans-Peter A. and Kvadsheim, Petter H. and Cur{\'e}, Charlotte},
  year = 2022,
  month = mar,
  journal = {Proceedings of the National Academy of Sciences},
  volume = {119},
  number = {13},
  pages = {e2114932119},
  publisher = {Proceedings of the National Academy of Sciences},
  doi = {10.1073/pnas.2114932119},
  urldate = {2022-04-07}
}

@article{seppanenSocialInformationUse2007,
  title = {Social {{Information Use Is}} a {{Process Across Time}}, {{Space}}, and {{Ecology}}, {{Reaching Heterospecifics}}},
  author = {Sepp{\"a}nen, Janne-Tuomas and Forsman, Jukka T. and M{\"o}nkk{\"o}nen, Mikko and Thomson, Robert L.},
  year = 2007,
  journal = {Ecology},
  volume = {88},
  number = {7},
  pages = {1622--1633},
  issn = {1939-9170},
  doi = {10.1890/06-1757.1},
  urldate = {2026-02-13},
  copyright = {\copyright{} 2007 by the Ecological Society of America},
  langid = {english}
}

@article{rendall2000proximate,
  title={Proximate factors mediating ``contact'' calls in adult female baboons (\textit{Papio cynocephalus ursinus}) and their infants.},
  author={Rendall, Drew and Cheney, Dorothy L and Seyfarth, Robert M},
  journal={Journal of Comparative Psychology},
  volume={114},
  number={1},
  pages={36},
  year={2000},
  publisher={American Psychological Association}
}

@article{branchMostFlightBaleen2025,
  title = {Most ``Flight'' Baleen Whale Species Are Acoustically Cryptic to Killer Whales, Unlike ``Fight'' Species},
  author = {Branch, Trevor A.},
  year = 2025,
  journal = {Marine Mammal Science},
  volume = {41},
  number = {3},
  pages = {e13228},
  issn = {1748-7692},
  doi = {10.1111/mms.13228},
  urldate = {2026-02-13},
  copyright = {\copyright{} 2025 Society for Marine Mammalogy.},
  langid = {english}
}

@article{morisakaPredationKillerWhales2007,
  title = {Predation by Killer Whales ({{Orcinus}} Orca) and the Evolution of Whistle Loss and Narrow-band High Frequency Clicks in Odontocetes},
  author = {Morisaka, T. and Connor, R. C.},
  year = 2007,
  month = jul,
  journal = {Journal of Evolutionary Biology},
  volume = {20},
  number = {4},
  pages = {1439--1458},
  issn = {1010-061X},
  doi = {10.1111/j.1420-9101.2007.01336.x},
  urldate = {2024-11-21}
}

@article{duarteSoundscapeAnthropoceneOcean2021a,
  title = {The Soundscape of the {{Anthropocene}} Ocean},
  author = {Duarte, Carlos M. and Chapuis, Lucille and Collin, Shaun P. and Costa, Daniel P. and Devassy, Reny P. and Eguiluz, Victor M. and Erbe, Christine and Gordon, Timothy A. C. and Halpern, Benjamin S. and Harding, Harry R. and Havlik, Michelle N. and Meekan, Mark and Merchant, Nathan D. and {Miksis-Olds}, Jennifer L. and Parsons, Miles and Predragovic, Milica and Radford, Andrew N. and Radford, Craig A. and Simpson, Stephen D. and Slabbekoorn, Hans and Staaterman, Erica and Van Opzeeland, Ilse C. and Winderen, Jana and Zhang, Xiangliang and Juanes, Francis},
  year = 2021,
  month = feb,
  journal = {Science},
  volume = {371},
  number = {6529},
  pages = {eaba4658},
  publisher = {American Association for the Advancement of Science},
  doi = {10.1126/science.aba4658},
  urldate = {2021-09-16}
}

@article{goldbogenIntegrativeApproachesStudy2013,
  title = {Integrative {{Approaches}} to the {{Study}} of {{Baleen Whale Diving Behavior}}, {{Feeding Performance}}, and {{Foraging Ecology}}},
  author = {Goldbogen, Jeremy A. and Friedlaender, Ari S. and Calambokidis, John and McKenna, Megan F. and Simon, Malene and Nowacek, Douglas P.},
  year = 2013,
  month = feb,
  journal = {BioScience},
  volume = {63},
  number = {2},
  pages = {90--100},
  issn = {0006-3568},
  doi = {10.1525/bio.2013.63.2.5},
  urldate = {2023-09-17}
}

@article{dunlopEavesdroppingRiskConspicuous2021,
  title = {The Eavesdropping Risk of Conspicuous Sexual Signaling in Humpback Whales},
  author = {Dunlop, Rebecca A. and Noad, Michael J.},
  year = 2021,
  month = aug,
  journal = {Behavioral Ecology and Sociobiology},
  volume = {75},
  number = {8},
  pages = {124},
  issn = {1432-0762},
  doi = {10.1007/s00265-021-03048-7},
  urldate = {2026-02-12},
  langid = {english}
}

@misc{dclde2013,
  author = {DCL/DE},
  title = {Baleen Whale Annotations from the Detection Classification Localization and Density Estimate Conference in 2013},
  year = {2013},
  howpublished = {NOAA National Centers for Environmental Information.},
  doi = {https://doi.org/10.25921/zaea-1s39},
  note = {Accessed: 2026-02-14}
}

@article{townsendAllClearMeerkats2011,
  title = {All Clear? {{Meerkats}} Attend to Contextual Information in Close Calls to Coordinate Vigilance},
  shorttitle = {All Clear?},
  author = {Townsend, Simon W. and Z{\"o}ttl, Markus and Manser, Marta B.},
  year = 2011,
  month = oct,
  journal = {Behavioral Ecology and Sociobiology},
  volume = {65},
  number = {10},
  pages = {1927--1934},
  issn = {1432-0762},
  doi = {10.1007/s00265-011-1202-6},
  urldate = {2025-01-28},
  langid = {english}
}

@article{manserInformationThatReceivers2001,
  title = {The Information That Receivers Extract from Alarm Calls in Suricates},
  author = {Manser, Marta B. and Bell, Matthew B. and Fletcher, Lindsay B.},
  year = 2001,
  month = dec,
  journal = {Proceedings of the Royal Society of London. Series B: Biological Sciences},
  volume = {268},
  number = {1484},
  pages = {2485--2491},
  issn = {0962-8452, 1471-2954},
  doi = {10.1098/rspb.2001.1772},
  urldate = {2025-02-07},
  langid = {english}
}

@incollection{manserChapterSixVocal2014,
  title = {Chapter {{Six}} - {{Vocal Complexity}} in {{Meerkats}} and {{Other Mongoose Species}}},
  booktitle = {Advances in the {{Study}} of {{Behavior}}},
  author = {Manser, Marta B. and Jansen, David A. W. A. M. and Graw, Beke and Holl{\'e}n, Linda I. and Bousquet, Christophe A. H. and Furrer, Roman D. and {le Roux}, Aliza},
  editor = {Naguib, Marc and Barrett, Louise and Brockmann, H. Jane and Healy, Sue and Mitani, John C. and Roper, Timothy J. and Simmons, Leigh W.},
  year = 2014,
  month = jan,
  volume = {46},
  pages = {281--310},
  publisher = {Academic Press},
  doi = {10.1016/B978-0-12-800286-5.00006-7},
  urldate = {2025-01-28}
}

@incollection{tyackCommunicationAcousticBehavior2000,
  title = {Communication and {{Acoustic Behavior}} of {{Dolphins}} and {{Whales}}},
  booktitle = {Hearing by {{Whales}} and {{Dolphins}}},
  author = {Tyack, Peter L. and Clark, Christopher W.},
  editor = {Au, Whitlow W. L. and Fay, Richard R. and Popper, Arthur N.},
  year = 2000,
  pages = {156--224},
  publisher = {Springer},
  address = {New York, NY},
  doi = {10.1007/978-1-4612-1150-1_4},
  urldate = {2026-02-10},
  isbn = {978-1-4612-1150-1},
  langid = {english}
}

@article{seyfarthSignalersReceiversAnimal2003,
  title = {Signalers and {{Receivers}} in {{Animal Communication}}},
  author = {Seyfarth, Robert M. and Cheney, Dorothy L.},
  year = 2003,
  month = feb,
  journal = {Annual Review of Psychology},
  volume = {54},
  number = {Volume 54, 2003},
  pages = {145--173},
  publisher = {Annual Reviews},
  issn = {0066-4308, 1545-2085},
  doi = {10.1146/annurev.psych.54.101601.145121},
  urldate = {2026-02-09},
  langid = {english}
}

@article{manserSuricateAlarmCalls2002,
  title = {Suricate Alarm Calls Signal Predator Class and Urgency},
  author = {Manser, Marta B. and Seyfarth, Robert M. and Cheney, Dorothy L.},
  year = 2002,
  month = feb,
  journal = {Trends in Cognitive Sciences},
  volume = {6},
  number = {2},
  pages = {55--57},
  publisher = {Elsevier},
  issn = {1364-6613, 1879-307X},
  doi = {10.1016/S1364-6613(00)01840-4},
  urldate = {2025-01-28},
  langid = {english}
}

@article{paynemcvay1971,
  title = {Songs of {{Humpback Whales}}},
  author = {Payne, Roger S. and McVay, Scott},
  year = 1971,
  month = aug,
  journal = {Science},
  volume = {173},
  number = {3997},
  pages = {585--597},
  publisher = {American Association for the Advancement of Science},
  doi = {10.1126/science.173.3997.585},
  urldate = {2026-02-13}
}

@article{Ross2021,
   author = {Gordon J Ross},
   doi = {10.1785/0120200198},
   isbn = {0120200198},
   issue = {3},
   journal = {Bulletin of the Seismological Society of America},
   pages = {1473-1480},
   title = {{Bayesian estimation of the ETAS model for earthquake occurrences}},
   volume = {111},
   url = {https://orcid.org/0000-0003-2092-0106},
   year = {2021}
}

@article{rasmussen2013bayesian,
  title={{Bayesian inference for Hawkes processes}},
  author={Rasmussen, Jakob Gulddahl},
  journal={Methodology and Computing in Applied Probability},
  volume={15},
  number={3},
  pages={623--642},
  year={2013},
  publisher={Springer}
}

@inproceedings{reynaud2013inference,
  title={{Inference of functional connectivity in neurosciences via Hawkes processes}},
  author={Reynaud-Bouret, Patricia and Rivoirard, Vincent and Tuleau-Malot, Christine},
  booktitle={2013 IEEE global conference on signal and information processing},
  pages={317--320},
  year={2013},
  organization={IEEE}
}

@inproceedings{zhou2013learning,
  title={{Learning social infectivity in sparse low-rank networks using multi-dimensional Hawkes processes}},
  author={Zhou, Ke and Zha, Hongyuan and Song, Le},
  booktitle={Artificial intelligence and statistics},
  pages={641--649},
  year={2013},
  organization={PMLR}
}

@article{embrechts2011multivariate,
  title={{Multivariate Hawkes processes: An application to financial data}},
  author={Embrechts, Paul and Liniger, Thomas and Lin, Lu},
  journal={Journal of Applied Probability},
  volume={48},
  number={A},
  pages={367--378},
  year={2011},
  publisher={Cambridge University Press}
}

@article{mohler2011self,
  title={{Self-exciting point process modeling of crime}},
  author={Mohler, George O and Short, Martin B and Brantingham, P Jeffrey and Schoenberg, Frederic Paik and Tita, George E},
  journal={Journal of the american statistical association},
  volume={106},
  number={493},
  pages={100--108},
  year={2011},
  publisher={Taylor \& Francis}
}

@incollection{rizoiu2017hawkes,
  title={Hawkes processes for events in social media},
  author={Rizoiu, Marian-Andrei and Lee, Young and Mishra, Swapnil and Xie, Lexing},
  booktitle={Frontiers of multimedia research},
  pages={191--218},
  year={2017},
  publisher={Association for Computing Machinery and Morgan \& Claypool}
}

@article{Ogata1998,
   author = {Yosihiko Ogata},
   doi = {10.1023/A:1003403601725/METRICS},
   issn = {00203157},
   issue = {2},
   journal = {Annals of the Institute of Statistical Mathematics},
   pages = {379-402},
   publisher = {Springer Netherlands},
   title = {{Space-time point-process models for earthquake occurrences}},
   volume = {50},
   year = {1998}
}

@article{Meyer2012,
   author = {Sebastian Meyer and Johannes Elias and Michael Höhle},
   doi = {10.1111/J.1541-0420.2011.01684.X},
   issn = {1541-0420},
   issue = {2},
   journal = {Biometrics},
   month = {6},
   pages = {607-616},
   pmid = {21981412},
   publisher = {John Wiley \& Sons, Ltd},
   title = {{A space–time conditional intensity model for invasive meningococcal disease occurrence}},
   volume = {68},
   year = {2012}
}

@article{Nicvert2024,
   author = {Lisa Nicvert and Sophie Donnet and Mark Keith and Mike Peel and Michael J. Somers and Lourens H. Swanepoel and Jan Venter and Hervé Fritz and Stéphane Dray},
   doi = {10.1002/ecy.4237},
   issn = {0012-9658},
   journal = {Ecology},
   month = {2},
   title = {{Using the multivariate Hawkes process to study interactions between multiple species from camera trap data}},
   volume = {e4237.},
   year = {2024}
}

@article{hawkes2018hawkes,
  title={{Hawkes processes and their applications to finance: a review}},
  author={Hawkes, Alan G},
  journal={Quantitative Finance},
  volume={18},
  number={2},
  pages={193--198},
  year={2018},
  publisher={Taylor \& Francis}
}

@article{Hawkes1971,
   author = {Alan G. Hawkes},
   issue = {1},
   journal = {Biometrika},
   pages = {83-90},
   title = {{Spectra of some self-exciting and mutually exciting point processes}},
   volume = {58},
   year = {1971},
}

@article{Lewis1979,
   author = {P. A.W. Lewis and G. S. Shedler},
   issue = {3},
   journal = {Naval research logistics quarterly},
   pages = {403-413},
   title = {{Simulation of nonhomogeneous Poisson processes by thinning}},
   volume = {26},
   year = {1979},
}

@article{Veen2008,
   author = {Alejandro Veen and Frederic P. Schoenberg},
   issue = {482},
   journal = {Journal of the American Statistical Association},
   pages = {614-624},
   title = {{Estimation of space-time branching process models in Seismology using an EM–type algorithm}},
   volume = {103},
   year = {2008},
}

@article{Sulem2024,
   author = {Déborah Sulem and Vincent Rivoirard and Judith Rousseau},
   doi = {10.3150/23-BEJ1631},
   issn = {13507265},
   issue = {2},
   journal = {Bernoulli},
   month = {5},
   pages = {1257-1286},
   title = {{Bayesian estimation of nonlinear Hawkes processes}},
   volume = {30},
   year = {2024}
}

@article{Mei2017,
   author = {Hongyuan Mei and Jason Eisner},
   journal = {Advances in neural information processing systems},
   title = {{The neural Hawkes process: A neurally self-modulating multivariate point process}},
   volume = {30},
   year = {2017}
}

@article{Bonnet2023,
   author = {Anna Bonnet and Miguel Martinez Herrera and Maxime Sangnier},
   doi = {10.1007/s11222-023-10264-w},
   issue = {91},
   journal = {Statistics and Computing},
   title = {{Inference of multivariate exponential Hawkes processes with inhibition and application to neuronal activity}},
   volume = {33},
   year = {2023}
}

@article{White2021,
   author = {Philip A. White and Alan E. Gelfand},
   issue = {3},
   journal = {Methodology and Computing in Applied Probability},
   pages = {1001-1021},
   title = {{Generalized evolutionary point processes: model specifications and model comparison}},
   volume = {23},
   year = {2021},
}

@article{Deutsch2025,
   author = {Isabella Deutsch and Gordon J. Ross},
   doi = {10.1214/24-AOAS1957},
   issue = {1},
   journal = {Annals of Applied Statistics},
   pages = {235-260},
   title = {{Estimating product cannibalisation in wholesale using multivariate Hawkes processes with inhibition}},
   volume = {19},
   year = {2025}
}

@inproceedings{Olinde2020,
   author = {Jack Olinde and Martin B. Short},
   doi = {10.1109/BigData50022.2020.9378017},
   isbn = {9781728162515},
   booktitle = {Proceedings - 2020 IEEE International Conference on Big Data, Big Data 2020},
   month = {12},
   pages = {3212-3219},
   publisher = {Institute of Electrical and Electronics Engineers Inc.},
   title = {A Self-limiting Hawkes Process: Interpretation, Estimation, and Use in Crime Modeling},
   year = {2020}
}

@article{Brown1988,
  title={A simple proof of the multivariate random time change theorem for point processes},
  author={Brown, Timothy C and Nair, M Gopalan},
  journal={Journal of Applied Probability},
  volume={25},
  number={1},
  pages={210--214},
  year={1988},
  publisher={Cambridge University Press}
}

@article{Demartsev2024,
   author = {Vlad Demartsev and Baptiste Averly and Lily Johnson-Ulrich and Vivek H. Sridhar and Leonardos Leonardos and Alexander Vining and Mara Thomas and Marta B. Manser and Ariana Strandburg-Peshkin},
   doi = {10.1098/rstb.2023.0188},
   issn = {14712970},
   issue = {1905},
   journal = {Philosophical Transactions of the Royal Society B: Biological Sciences},
   pmid = {38768207},
   publisher = {Royal Society Publishing},
   title = {{Mapping vocal interactions in space and time differentiates signal broadcast versus signal exchange in meerkat groups}},
   volume = {379},
   year = {2024}
}

@article{Kang2025whaleHawkes,
author = {Bokgyeong Kang and Erin M. Schliep and Alan E. Gelfand and Tina M. Yack and Christopher W. Clark and Robert S. Schick},
title = {{Analyzing whale calling through Hawkes process modeling}},
journal = {Journal of the American Statistical Association},
volume = {120},
number = {552},
pages = {2040--2052},
year = {2025},
publisher = {Taylor \& Francis},
doi = {10.1080/01621459.2025.2501711}
}

@article{watanabe2013widely,
  title={{A widely applicable Bayesian information criterion}},
  author={Watanabe, Sumio},
  journal={The Journal of Machine Learning Research},
  volume={14},
  number={1},
  pages={867--897},
  year={2013},
  publisher={JMLR. org}
}

\end{document}


\def\spacingset#1{\renewcommand{\baselinestretch}%
{#1}\small\normalsize} \spacingset{1}


\if1\blind
{
  \title{\bf Supplementary Material for ``Modeling Animal Communication Using Multivariate Hawkes Processes with Additive Excitation and Multiplicative Inhibition''}
  \author{Bokgyeong Kang, Erin M. Schliep, Alan E. Gelfand, \\
  Ariana Strandburg-Peshkin, and Robert S. Schick
  }
  \maketitle
} \fi

\if0\blind
{
  \bigskip
  \bigskip
  \bigskip
  \begin{center}
    {\LARGE\bf Supplementary Material for ``??''}
\end{center}
  \medskip
} \fi

\spacingset{1.9} 



\section{Identifiability of the Self-Limiting Hawkes Process}
\label{sup:sec:olinde}

This section examines the identifiability of the excitation and inhibition
parameters, $\alpha$ and $\gamma$, in the self-limiting Hawkes process of
\citet{Olinde2020} under settings where both effects are pronounced.
For $t \in (0, 1000]$, point process realizations are generated from the model
using the parameter values
\begin{align*}
\mu = e^{\beta} = 0.65, \quad
\alpha = 0.65, \quad
\gamma = 0.3, \quad
\eta = 5, \quad
\phi = 3.
\end{align*}
These values are selected to induce substantial excitation and inhibition.
Each realization is then fitted using the same model as that employed for data
generation, so that the generating and fitting models coincide.

\begin{figure}[H]
    \centering
    \includegraphics[width=\linewidth]{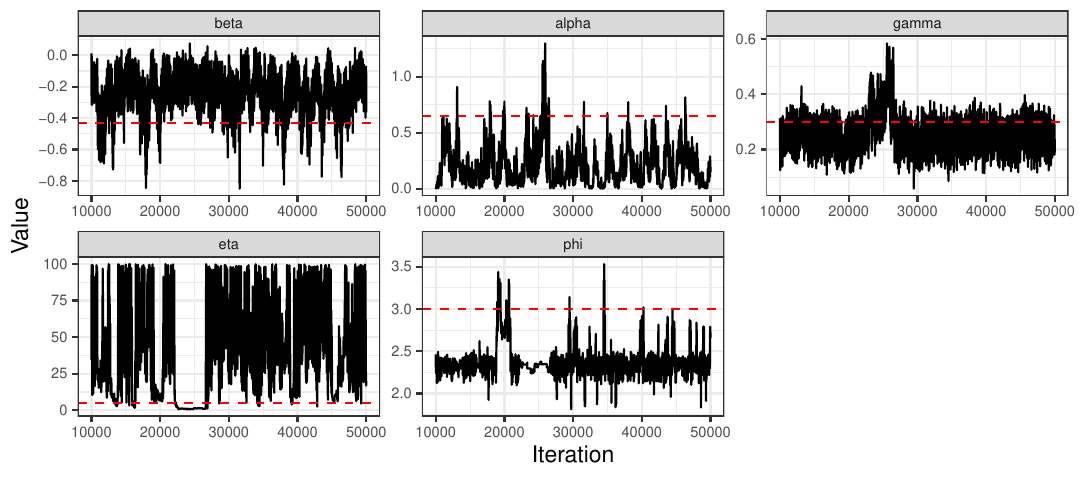}
    \caption{Trace plots of posterior samples for the parameters of the self-limiting Hawkes process of Olinde and Short (2020). Dashed red lines indicate the true parameter values.}
    \label{sup:fig:olinde}
\end{figure}

Figure~\ref{sup:fig:olinde} displays the trace plots of posterior samples for the
model parameters. The results indicate potential identifiability challenges in this setting. In particular, the posterior samples do not concentrate around the true value
of the excitation parameter, suggesting that the excitation effect is
systematically underestimated. This appears to be partially offset by a reduction in the inferred inhibition effect, most notably through smaller values of $\phi$, which effectively limit the number of past events contributing to suppression.

We note that \citet{Olinde2020} focus on regimes in which suppression is present but relatively weak, considering values of $\gamma$ in the range $\gamma \in [0.005, 0.05]$ in their simulation studies. The present results suggest that, when the suppression effect is stronger, the excitation and inhibition parameters in the self-limiting Hawkes model may become difficult to disentangle.

\bibliographystyle{apalike}
\bibliography{refs}